\documentclass[aps,prc,showkeys,nofootinbib,preprintnumbers]{revtex4}

\usepackage{color} 

\usepackage{graphicx}

\usepackage{isotope}
\usepackage{physics}

\newcommand{\sigmabf}{\mbox{\boldmath $\sigma$}}

\newcommand{\rhobf}{\mbox{\boldmath $\rho$}}

\usepackage[normalem]{ulem}

\def\SCNTHuizhou{$^{(1)}$Southern Center for Nuclear-Science Theory (SCNT), Institute of Modern Physics, Chinese Academy of Sciences, Huizhou 516000, China}
\def\kyiv{$^{(2)}$Institute for Nuclear Research, National Academy of Sciences of Ukraine, Kyiv, 03680, Ukraine}

\def\ucsaopaolo{
$^{(3)}$Laborat\'{o}rio de F\'{i}sica Te\'{o}rica e Computacional - LFTC,
Programa de P\'{o}sgradua\c{c}\~{a}o em Astrof\'{i}sica e F\'{i}sica Computacional,
Universidade Cidade de S\~{a}o Paulo, 01506-000 S\~{a}o Paulo, S\~{a}o Paulo, Brazil
}

\def\omegisuseoul{$^{(4)}$Department of Physics and OMEG Institute, Soongsil University,
Seoul 06978, Republic of Korea}

\def\sunyatsen{$^{(5)}$School of Physics and Astronomy, Sun Yat-Sen University, Zhuhai, 519082, China}

\begin{document}

\phantom{0}    
\vspace{-0.2in}  
\hspace{5.5in}

\preprint{{\bf LFTC-25-03/97}}

\title{%
Manifestation 
of quark effects in nuclei via bremsstrahlung analysis in the proton-nucleus scattering}
\author{Sergei~P.~Maydanyuk$^{(1,2)}$}\email{sergei.maydanyuk@impcas.ac.cn}%
\author{K.~Tsushima$^{(3)}$}\email{kazuo.tsushima@gmail.com}%
\author{G.~Ramalho$^{(4)}$}\email{gilberto.ramalho2013@gmail.com}%
\author{Peng-Ming~Zhang$^{(5)}$}\email{zhangpm5@mail.sysu.edu.cn}%

\affiliation{\SCNTHuizhou}
\affiliation{\kyiv}
\affiliation{\ucsaopaolo}
\affiliation{\omegisuseoul}
\affiliation{\sunyatsen}


\date{\small\today}

\begin{abstract}
\begin{description}
\item[Background]
(1) The incoherent emission of photons
is known to be dominant comparing to the coherent one
in proton-nucleus scattering.
The incoherent bremsstrahlung is
very sensitive to the magnetic moments of nucleons in nuclei.
(2) According to the quark-meson coupling (QMC) model,
the nucleon magnetic moments in nuclei are enhanced
relative to those in vacuum, originating from the quark structure of nucleons.
%
%
\item[Purpose]
Investigate possibilities of observing quark effects in nuclei
by the analysis of bremsstrahlung photons in nuclear reactions.
%
\item[Methods]
Analyse the bremsstrahlung emission differential cross sections
with the existing established bremsstrahlung model in proton-nucleus
scattering, by extending with the inclusion of in-medium modified nucleon magnetic moments
in nuclei calculated by the QMC model.
%
\item[Results]
(1) After calibrating the model without the quark effects for the experimental data of
bremsstrahlung in proton-nucleus scattering
(TAPS Collaboration data for \isotope[197]{Au}
at proton beam energy $E_{\rm p} = 190$~MeV),
we calculate the differential cross sections including the
in-medium enhanced nucleon magnetic moments due to the quark effects in the
\isotope[197]{Au} nucleus.
We observe the slight difference between the spectra for the models
with and without the quark effects
in the wide energy range of photons.
Such result is found for the first time, confirming possibilities
of observing the quark effects in the spectra of bremsstrahlung emission
in proton-nucleus scattering.
%
(2) We proceed to find the nucleus which has similar 
amount of contributions for the incoherent and coherent bremsstrahlung, to make 
more suitable for observing the quark effects in the bremsstrahlung.
It turns out that the quark effects are not sufficiently enough to be able to observe
clearly in heavy nuclei, since they have dominant incoherent contribution.
Then, we proceed to study from \isotope[197]{Au} to nuclei of middle masses.
For \isotope[40]{Ca} we find that the
coherent and incoherent contributions are similar at the low photon-energy region
(up to 0.4~MeV). Thus, we conclude that this nucleus is still not suitable to observe
the quark effects clearly.
%
(3) For \isotope[16]{O} we expect to see an increase of the coherent contribution.
Our estimates show that visibility of the quark effects for \isotope[16]{O}
is better than \isotope[40]{Ca},
although the coherent and incoherent contributions are similar at the
photon energy of 5~MeV.
%
(4)
By analyzing carbon isotopes, we find that \isotope[18]{C} has
minimal incoherent contribution,
where the quark effects should be minimal in comparison with the other carbon isotopes.
Ratios between the spectra for \isotope[18]{C} and those
for \isotope[12]{C} with and without the quark effects,
the difference is clearly observed.
Thus, the ratios of these spectra can provide the most evident information on
the quark effects in nuclei.
%
\item[Conclusions]
We establish
the new physical observable
for the quark effects in nuclei
in the bremsstrahlung photon emission
accompanied in the nuclear reactions,
which can be measured in experiments.
The present suggestion is for the first time
in both theoretically and experimentally
to study the quark effects in nuclei
via the bremsstrahlung.
\end{description}

\end{abstract}



\maketitle

\emph{Introduction.
\label{sec.introduction}}\;
By the studies of bremsstrahlung in scattering of protons off nuclei,
the incoherent photon emission is known to give the
dominant contribution than that of the coherent one
(see Ref.~\cite{Maydanyuk.2023.PRC.delta}, and reference therein).
In particular, according to the estimates in Ref.~\cite{Maydanyuk_Zhang.2015.PRC},
the contribution from the incoherent bremsstrahlung is about $10^{+6}$--$10^{+7}$
times larger than that of the coherent one for the nuclei studied.
So far, it is confirmed as the most accurate agreement of the calculated cross sections,
practically for the whole energy region of the emitted photons with the
experimental data~\cite{Goethem.2002.PRL} for $p + \isotope[12]{C}$,
$p + \isotope[58]{Ni}$, $p + \isotope[107]{Ag}$ and $p + \isotope[197]{Au}$ at $E_{\rm p}=190$~MeV
(see Figs.~6, 7 in Ref.~\cite{Maydanyuk_Zhang.2015.PRC}), and
experimental data~\cite{Clayton.1991.PhD,Clayton.1992.PRC} for
$p + \isotope[208]{Pb}$ at proton energy beam of $E_{\rm p}=145$~MeV
(see Figs.~5 in Ref.~\cite{Maydanyuk_Zhang.2015.PRC}).

The dominance of the incoherent contribution was studied in a more systematic way
from the light to heavy nuclei in the wide proton beam energy region,
and confirmed as a general phenomenon of bremsstrahlung in the proton-nucleus
scattering~\cite{Maydanyuk_Zhang.2015.PRC},
with possible exceptions of deuteron and some very light nuclei.
This holds also for nuclei with $\Delta$ baryons from \isotope[12][\Delta]{C} to
\isotope[208][\Delta]{Pb}
in the proton-nucleus scattering~\cite{Maydanyuk.2023.PRC.delta}.
Furthermore, the important role of the incoherent bremsstrahlung in reactions
with hypernuclei is also confirmed~\cite{Liu_Maydanyuk_Zhang_Liu.2019.PRC.hypernuclei}
(see predictions in the paper for
\isotope[10][\Lambda]{Be}, \isotope[10][\Lambda]{B},
\isotope[107][\Lambda]{Te}, \isotope[109][\Lambda]{Te}, \isotope[211][\Lambda]{Po}
in comparison with \isotope[9]{Be}, \isotope[9]{B}, \isotope[106]{Te}, \isotope[108]{Te},
\isotope[210]{Po}).
The photon energy dependence of the differential cross section of the incoherent bremsstrahlung
contribution has a plateau,
while the coherent bremsstrahlung one has a logarithmic type dependence.
This property allows one to distinguish the two contributions in the full cross section
(via the ratio between the incoherent and coherent contributions)
using measured data of bremsstrahlung.
(For example, see data~\cite{Goethem.2002.PRL} of TAPS Collaboration and
analysis~\cite{Maydanyuk_Zhang.2015.PRC}.)

It turns out that the magnetic moments of nucleons in nuclei (and in the proton beam)
play key role in estimating the incoherent contribution,
while this feature is nearly absent in the coherent bremsstrahlung.
Therefore, any small variations in the nucleon magnetic moments
in proton-nucleus scattering can give visible changes in the full calculated cross section of
bremsstrahlung.
Note that, the full cross section can be tested by measuring the emitted photons.
Thus, the key interest is the role of the nucleon magnetic moments in nuclei.

On the other side, the quark-meson coupling (QMC) model predicts that
the magnetic moments of nucleons in nuclei are enhanced relative to
those in free
space~\cite{Guichon:1987jp,Saito:2005rv,Krein:2017usp,Guichon:2018uew,Lu:1997mu,Lu:1998tn}.
Thus, we can expect that such changes in the nucleon magnetic moments can give visible differences in
the calculated cross sections of bremsstrahlung.
Namely, new possible information on the quark effects in nuclei 
can be extracted from experimental bremsstrahlung cross sections. 

Such aspects of questions have never been addressed in the past.
In the present study, the QMC model
\cite{Guichon:1987jp,Saito:2005rv,Krein:2017usp,Guichon:2018uew,Lu:1997mu,Lu:1998tn}
is expected to be able to provide a suitable description of the quark effects in nuclear reactions.
It is a natural practice to combine the predictions of the QMC model with the established bremsstrahlung formalism
(see Refs.~\cite{Maydanyuk.2023.PRC.delta}, and reference therein),
in order to explore possible quark effects in nuclei using well tested reactions.
This is one of the main aims of this article.
In this article we report on the new theoretical results for such questions, 
focusing on the quark effects in nuclei, by analyzing the accompanied emission of bremsstrahlung
photons.
This is the first attempt to explore the possible quark effects in nuclei
by studying the bremsstrahlung photon emission in nuclear reactions.


\vspace{1.5mm}
\emph{Calculation of bremsstrahlung emission cross section in proton-nucleus scattering.
\label{sec.2.1}}\;
Formalism for describing the emission of bremsstrahlung photons without
the in-medium modified nucleon magnetic moments was presented in Ref.~\cite{Maydanyuk.2023.PRC.delta}.
Here, we give briefly the main ideas and formulas 
(see Supplemental Material~\cite{Maydanyuk.2025.Supplemental} for details).
Hamiltonian for scattering of a proton on a nucleus with the inclusion of emitting
photons in the laboratory frame is written with many-nucleon generalization of Pauli equation as
$\hat{H} = \hat{H}_{0} + \hat{H}_{\gamma}$.
Here, $\hat{H}_{0}$ is the Hamiltonian describing the nucleons in the
nucleus and the proton in the beam in the scattering
without the emission of photons, where, $\hat{H}_{\gamma}$ is operator describing
the emission of bremsstrahlung photons in the scattering,
and the terms are defined in Sec.~I.A of Supplemental Material~\cite{Maydanyuk.2025.Supplemental}.

For the analysis of the emission of photons, we need relative 
coordinate for the electric charges and magnetic moments of nucleons
in proton-nucleus scattering.
We rewrite formalism in relative coordinates and the corresponding relative momenta.
%
Operator for the emission of 
photons in the scattering of
proton on nucleus in the laboratory frame is written as
$\hat{H}_{\gamma} = \hat{H}_{P} + \hat{H}_{p}
+ \Delta \hat{H}_{\gamma E} + \Delta \hat{H}_{\gamma M} + \hat{H}_{k}$.
Here, 
$\hat{H}_{P}$ is the term related with the motion of the full proton-nucleus system,
while $\hat{H}_{p}$ is the term of the coherent emission,
$\Delta \hat{H}_{\gamma E}$ and $\Delta \hat{H}_{\gamma M}$ are the terms
of the incoherent emission of electric and magnetic types, 
and $\hat{H}_{k}$ is the background term.
These terms are defined in Sec.~I.C of Supplemental Material~\cite{Maydanyuk.2025.Supplemental}.

We define the matrix element of photon emission, using the wave functions $\Psi_{i}$ and
$\Psi_{f}$ of the
full nuclear system before the emission of photons (initial, $i$-state)
and after the emission (final, $f$-state), respectively, as
$\langle \Psi_{f} |\, \hat{H}_{\gamma} |\, \Psi_{i} \rangle$.
Calculation of the full matrix element of bremsstrahlung emission is straightforward,
and we write down result as
(see Sec.~I.D in Supplemental Material~\cite{Maydanyuk.2025.Supplemental})
$M_{\rm full} = M_{P} + M_{p}^{(E)} + M_{p}^{(M)} + M_{k} + M_{\Delta E} + M_{\Delta M}$,
where
\begin{equation}
\begin{array}{lllll}
\vspace{-0.1mm}
  M_{P} & = &
  \displaystyle\frac{\hbar\, (2\pi)^{3}}{m_{A} + m_{p}}\, \mu_{0}\,
  \displaystyle\sum\limits_{\alpha=1,2}
  \displaystyle\int\limits_{}^{}
    \Phi_{\rm p - nucl, f}^{*} (\vb{r})\;
  \biggl\{
    2\, m_{\rm p}\;
    \Bigl[
      e^{-i\, c_{A}\, \vb{k_{\rm ph}} \vb{r}} F_{p,\, {\rm el}} + e^{i\, c_{p}\, \vb{k_{\rm ph}} \vb{r}} F_{A,\, {\rm el}}
    \Bigr]\, \bigl( \vb{e}^{(\alpha)} \cdot \vb{K}_{i} \bigr)\; 
  \\
%
  & + &
    i\: 
    \biggl( \Bigl[
      e^{-i\, c_{A}\, \vb{k_{\rm ph}} \vb{r}}\, \vb{F}_{p,\, {\rm mag}} + e^{i\, c_{p}\, \vb{k_{\rm ph}} \vb{r}}\, \vb{F}_{A,\, {\rm mag}}
    \Bigr] \cdot
    \bigl[ \vb{K}_{i} \cp \vb{e}^{(\alpha)} \bigr]
    \biggr)
  \biggr\}\;
  \Phi_{\rm p - nucl, i} (\vb{r})\; \vb{dr}, \\
\vspace{-0.2mm}
  M_{p}^{(E)} & = &
  2\,i \hbar\, (2\pi)^{3} \displaystyle\frac{m_{\rm p}}{\mu}\: \mu_{0}
  \displaystyle\sum\limits_{\alpha=1,2}
  \displaystyle\int\limits_{}^{}
    \Phi_{\rm p - nucl, f}^{*} (\vb{r})\;
    e^{-i\, \vb{k}_{\rm ph} \vb{r}}\;
    Z_{\rm eff} (\vb{k}_{\rm ph}, \vb{r}) \; 
    \Bigl( \vb{e}^{(\alpha)} \cdot \vb{\displaystyle\frac{d}{dr}}\; \Phi_{\rm p - nucl, i} (\vb{r}) \Bigr)\; \vb{dr}, \\
  M_{p}^{(M)} & = &
  -\, \hbar\, (2\pi)^{3} \displaystyle\frac{m_{\rm p}}{\mu}\: \mu_{0}
  \displaystyle\sum\limits_{\alpha=1,2}
  \displaystyle\int\limits_{}^{}
    \Phi_{\rm p - nucl, f}^{*} (\vb{r})\;
    e^{-i\, \vb{k}_{\rm ph} \vb{r}} \;
    \left\{ \vb{M}_{\rm eff} (\vb{k}_{\rm ph}, \vb{r}) \cdot \Bigl[ \vb{\displaystyle\frac{d}{dr}}
\times \vb{e}^{(\alpha)} \Bigr]\;
    \Phi_{\rm p - nucl, i} (\vb{r}) \right\} \; \vb{dr},
\end{array}
\label{eq.13.1.4.p-nucl}
\end{equation}
\begin{equation}
\begin{array}{lll}
  M_{\Delta E} & = &
  -\, (2\pi)^{3}\, 2\, \mu_{0}
  \displaystyle\sum\limits_{\alpha=1,2} 
  \displaystyle\int\limits_{}^{}
    \Phi_{\rm p - nucl, f}^{*} (\vb{r})\;
    e^{i\, c_{p}\, \vb{k_{\rm ph}} \vb{r}}\, 
  \biggl\{
    \Bigl( \vb{e}^{(\alpha)} \cdot \vb{D}_{A 1,\, {\rm el}} \Bigr) -
    \displaystyle\frac{m_{\rm p}}{m_{A}}\, 
     \Bigl( \vb{e}^{(\alpha)} \cdot \vb{D}_{A 2,\, {\rm el}} \Bigr)
  \biggr\}\;
  \Phi_{\rm p - nucl, i} (\vb{r})\; \vb{dr}, \\
  M_{\Delta M} & = &
  -\, i\, (2\pi)^{3}\,  \mu_{0}\,
  \displaystyle\sum\limits_{\alpha=1,2}
  \displaystyle\int\limits_{}^{}
    \Phi_{\rm p - nucl, f}^{*} (\vb{r})\;
  \biggl\{
    e^{i\, c_{p}\, \vb{k_{\rm ph}} \vb{r}}\; D_{A 1,\, {\rm mag}}^{(\alpha)} - 
    e^{i\, c_{p}\, \vb{k_{\rm ph}} \vb{r}}\; D_{A 2,\, {\rm mag}}^{(\alpha)} 
  \biggr\}\;
  \Phi_{\rm p - nucl, i} (\vb{r})\; \vb{dr},
\end{array}
\label{eq.13.1.6.p-nucl}
\end{equation}

\vspace{-5.0mm}
\begin{equation}
\begin{array}{lcl}
  M_{k} & = &
  i\, \hbar\, (2\pi)^{3}  \mu_{0}\,
  \displaystyle\sum\limits_{\alpha=1,2}
  \displaystyle\int\limits_{}^{}
    \Phi_{\rm p - nucl, f}^{*} (\vb{r})\;
    \Bigl(
      \bigl[ \vb{k_{\rm ph}} \cp \vb{e}^{(\alpha)} \bigr] \cdot
      \Bigl\{ e^{-i\, c_{A}\, \vb{k_{\rm ph}} \vb{r}}\, \vb{D}_{p,\, {\rm k}} + e^{i\, c_{p}\, \vb{k_{\rm ph}} \vb{r}}\, \vb{D}_{A,\, {\rm k}} \Bigr\}
    \Bigr)\;
    \Phi_{\rm p - nucl, i} (\vb{r})\; \vb{dr}.
\end{array}
\label{eq.13.1.7.p-nucl}
\end{equation}
Here, 
$\vb{K}_{i} = \vb{K}_{f} + \vb{k}_{\rm ph}$,
$\mu = m_{\rm p} m_{A} / (m_{\rm p} + m_{A})$ is the reduced mass,
$\mu_{0} = e\hbar / (2m_{\rm p}c)$ is the nuclear magneton,
$c_{A} = \frac{m_{A}}{m_{A}+m_{\rm p}}$,
$c_{\rm p} = \frac{m_{\rm p}}{m_{A} + m_{\rm p}}$,
where $m_{\rm p}$ is the proton mass,
and $m_{A}$ is the mass of the target nucleus with the nucleon number $A$.
$\Phi_{\rm p - nucl} (\vb{r})$ is the function describing the relative motion
between the scattered proton and the target nucleus in the
scattering~\cite{Maydanyuk.2025.Supplemental}.
The effective electric charge $Z_{\rm eff}$ and magnetic moment $\vb{M}_{\rm eff}$ are
(see Eqs.~(28) and (29) in Ref.~\cite{Maydanyuk.2023.PRC.delta.Supplemental}):
\begin{equation}
\begin{array}{lll}
\vspace{1.0mm}
  Z_{\rm eff} (\vb{k}_{\rm ph}, \vb{r}) =
  e^{i\, \vb{k_{\rm ph}} \vb{r}}\,
  \Bigl[
    e^{-i\, c_{A} \vb{k_{\rm ph}} \vb{r}}\, \displaystyle\frac{m_{A}}{m_{p} + m_{A}}\, F_{p,\, {\rm el}} -
    e^{i\, c_{p} \vb{k_{\rm ph}} \vb{r}}\, \displaystyle\frac{m_{p}}{m_{p} + m_{A}}\, F_{A,\, {\rm el}}
  \Bigr], \\
  \vb{M}_{\rm eff} (\vb{k}_{\rm ph}, \vb{r}) =
  e^{i\, \vb{k_{\rm ph}} \vb{r}}\,
  \Bigl[
    e^{-i\, c_{A} \vb{k_{\rm ph}} \vb{r}}\,  \displaystyle\frac{m_{A}}{m_{p} + m_{A}}\, \vb{F}_{p,\, {\rm mag}} -
    e^{i\, c_{p} \vb{k_{\rm ph}} \vb{r}}\,  \displaystyle\frac{m_{p}}{m_{p} + m_{A}}\, \vb{F}_{A,\, {\rm mag}}
  \Bigr].
\end{array}
\label{eq.13.1.8.p-nucl}
\end{equation}
Here,
$F_{{\rm p},\, {\rm el}}$,
$F_{A,\, {\rm el}}$,
$\vb{F}_{{\rm p},\, {\rm mag}}$,
$\vb{F}_{A,\, {\rm mag}}$,
$\vb{D}_{A 1,\, {\rm el}}$,
$\vb{D}_{A 2,\, {\rm el}}$,
$D_{A 1,\, {\rm mag}}$,
$D_{A 2,\, {\rm mag}}$,
$\vb{D}_{{\rm p},\, {\rm k}}$,
$\vb{D}_{A,\, {\rm k}}$,
$D_{{\rm p}, P\, {\rm el}}$,
$D_{A,P\, {\rm el}}$,
$\vb{D}_{{\rm p}, P\, {\rm mag}}$,
$\vb{D}_{A,P\, {\rm mag}}$
are electric and magnetic form factors calculated 
on the basis of nucleons in the nucleus
(see Sec.~II in Supplemental Material~\cite{Maydanyuk.2025.Supplemental}).
These form factors are dependent on the energy of photon emitted.
%
Calculation of form factors for the incoherent term $M_{\Delta M}$ gives

\vspace{-2.0mm}
\noindent 
\begin{equation}
\begin{array}{llllll}
  \displaystyle\sum\limits_{\alpha=1,2} D_{A 1,\, {\rm mag}}^{(\alpha)} =
  -\, \displaystyle\frac{\hbar\, (A-1)}{2\,A}\; \bar{\mu}_{\rm pn}^{\rm (A)}\, |{\bf k_{\rm ph}}| \; Z_{\rm A}, & 
\hspace{4ex}  
D_{A 2,\, {\rm mag}}^{(\alpha)} = 0.
\end{array}
\label{eq.app.form_factors.1.4}
\end{equation}

\vspace{-2.0mm}
\noindent 
In the above,
$\bar{\mu}_{\rm pn}^{\rm (A)} = \mu_{\rm p} + \kappa_{A}\,\mu_{\rm n}$,
$\kappa_{A} = N_{A}/Z_{A}$, and
$\mu_{\rm p}$ and $\mu_{\rm n}$ are respectively 
the magnetic moments of proton and neutron, where,
$A$, $Z_{A}$ and $N_{A}$ are respectively the numbers of nucleons, protons and neutrons of
the nucleus.
In Eq.~(\ref{eq.app.form_factors.1.4}) one can see the role of the nucleon
magnetic moments in the nucleus
(i.e., variations of nucleon magnetic moments can change the full bremsstrahlung cross section).
Influence of variations in $\vb{F}_{A,\, {\rm mag}}$ on the full bremsstrahlung
is smaller compared to the dominant incoherent emission.
%
%
%
We define the cross sections of the bremsstrahlung 
in the proton-nucleus scattering
based on the full matrix element in the frameworks of in
Refs.~\cite{Maydanyuk.2023.PRC.delta,Maydanyuk_Zhang_Zou.2016.PRC,Maydanyuk.2012.PRC,Maydanyuk_Zhang.2015.PRC},
and Sec.~I.E in Supplemental Material~\cite{Maydanyuk.2025.Supplemental}.

According to the QMC model, the nucleon magnetic moments in nuclei
should be enhanced due to the quark substructure of the nucleons.
So, we will include the effect of the enhanced nucleon magnetic moments in the nucleus, and
calculate the bremsstrahlung cross sections.
Note that
this attempt is for the first time associated with the nuclear bremsstrahlung.

\begin{figure}[htbp]
\centerline{\includegraphics[width=74mm]{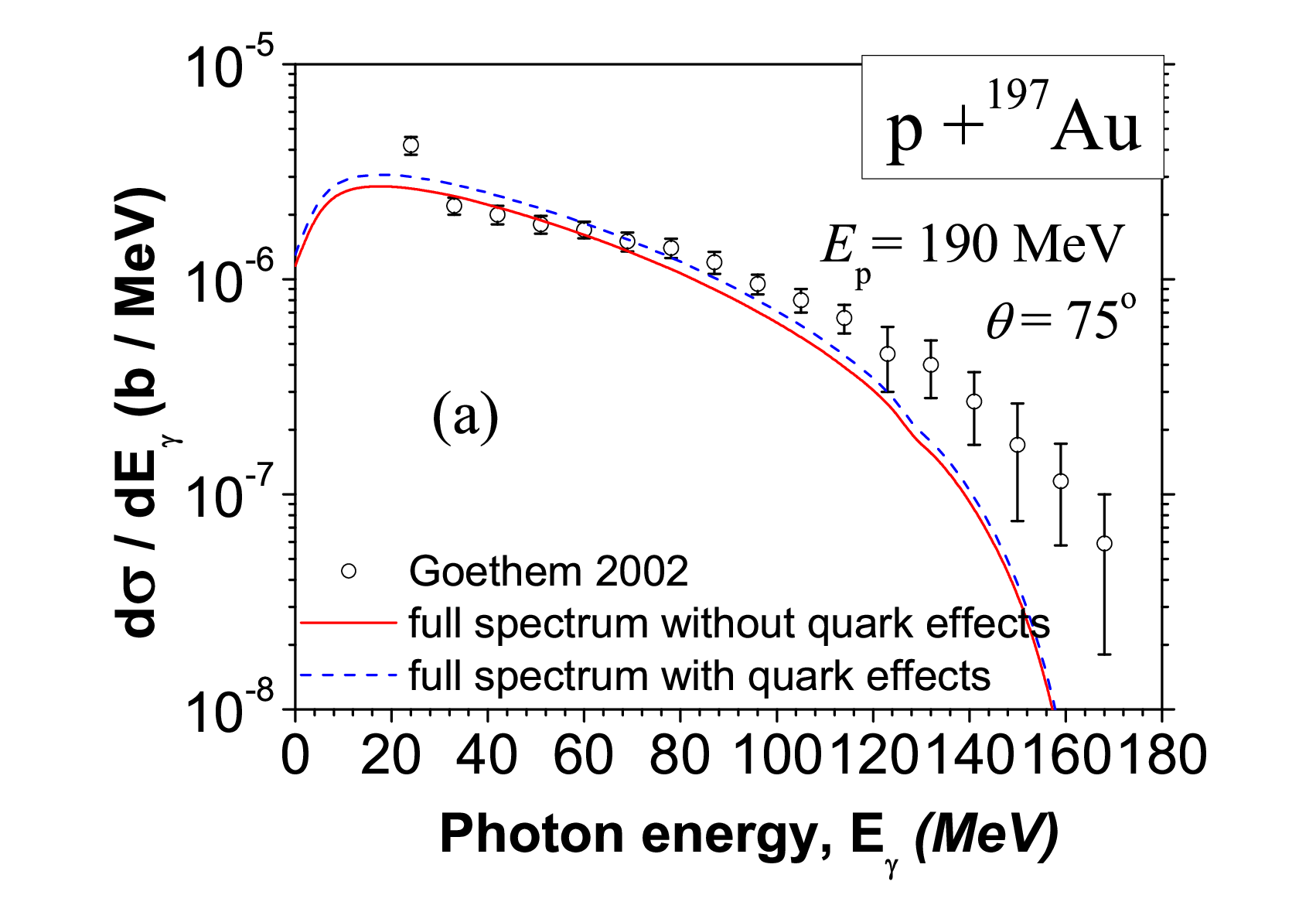}
\hspace{-13.5mm}\includegraphics[width=74mm]{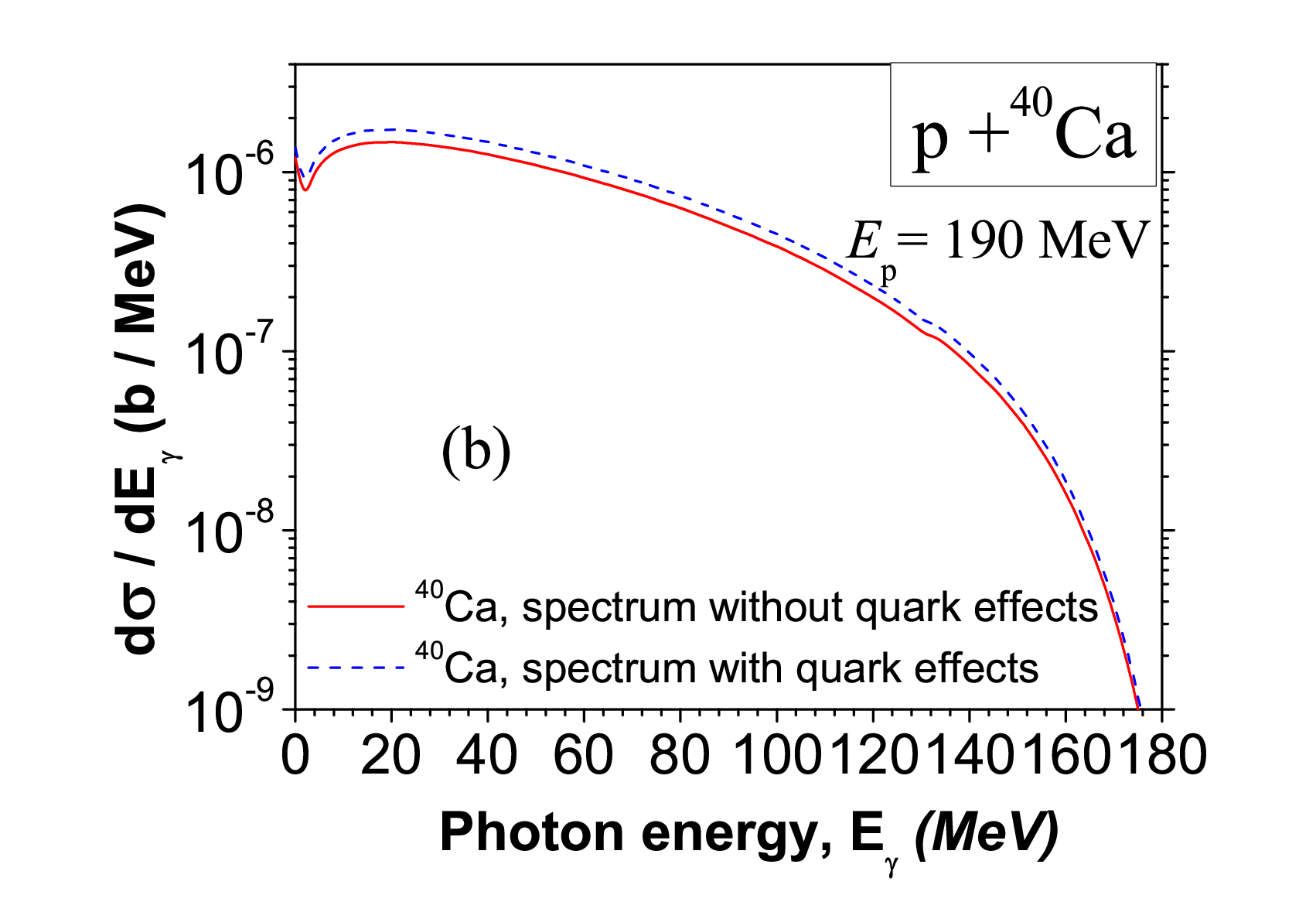}
\hspace{-13.5mm}\includegraphics[width=74mm]{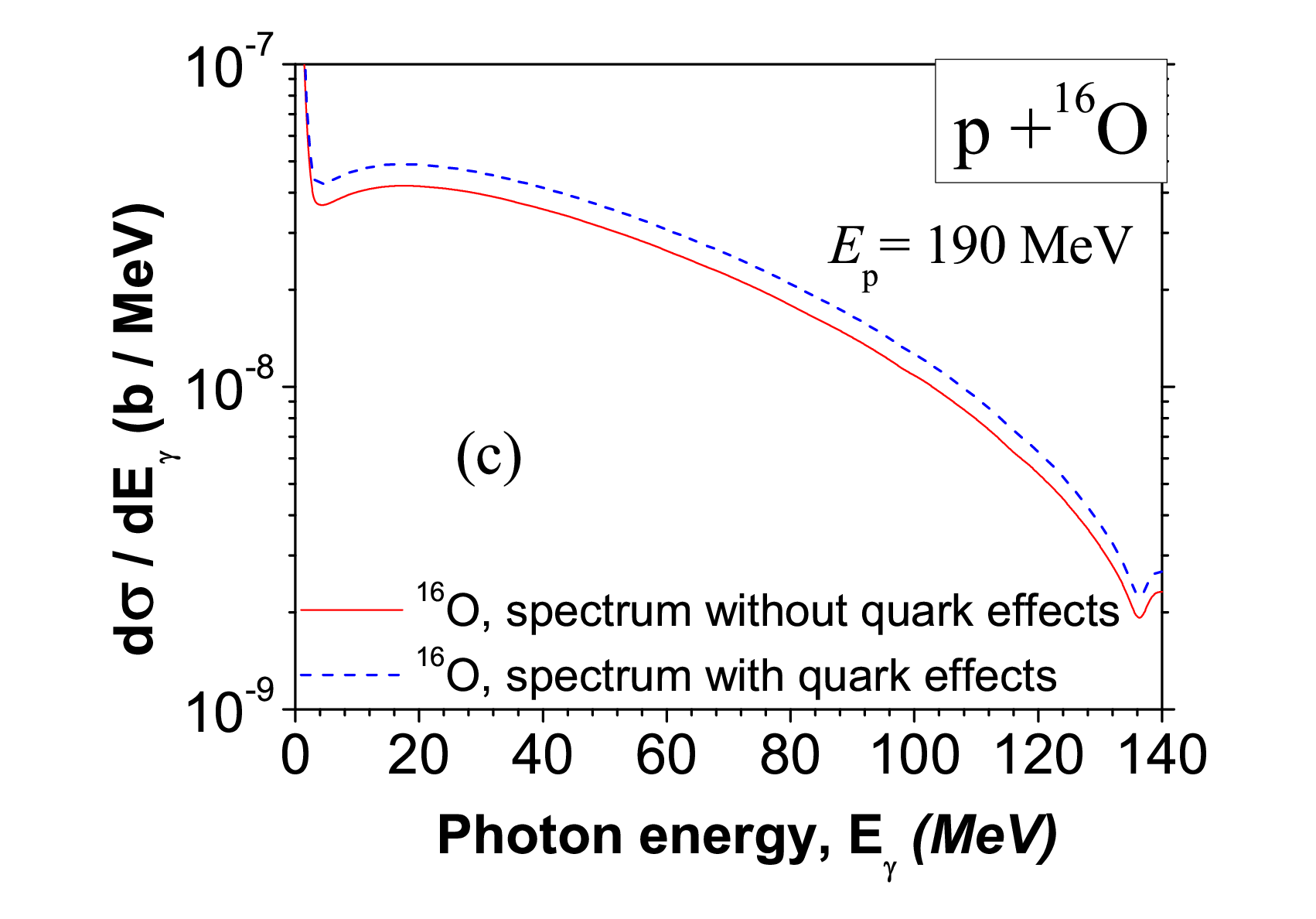}}
\vspace{-5mm}
\caption{\small (Color online)
Calculated bremsstrahlung spectra with the coherent and incoherent terms
in the scattering of protons off 
(a) nucleus \isotope[197]{Au},
(b) nucleus \isotope[40]{Ca}, and
(c) nucleus \isotope[16]{O},
at proton beam energy of $E_{\rm p}=190$~MeV
with and without the quark effects.
The experimental data for \isotope[197]{Au} given by open circles (Goethem 2002)
are extracted from
Ref.~\cite{Goethem.2002.PRL},
the red solid line is full spectrum without the quark effects,
while the blue dashed line is full spectrum with the quark effects.
All cross sections are calculated with the coherent plus incoherent contributions defined by
$M_{p}^{(E)}$, $M_{p}^{(M)}$, $M_{\Delta M}$ and $M_{k}$
in Eqs.~(\ref{eq.13.1.4.p-nucl})--(\ref{eq.13.1.7.p-nucl}).
\label{fig.analysis.197Au.1}}
\end{figure}

\vspace{1.5mm}
\emph{Sensitivity of bremsstrahlung cross sections on the quark effects
(enhanced nucleon magnetic moments in the nucleus).
Spectra of proton-nucleus scattering for \isotope[197]{Au}.
\label{sec.analysis.197Au}}\;
We first start by testing the model and algorithms of calculation of the bremsstrahlung
cross section for the \isotope[197]{Au} nucleus without the quark effects
at the proton beam energy $E_{\rm p} = 190$~MeV,
and compare with the experimental data~\cite{Goethem.2002.PRL}.
By this, the present model is calibrated using
the observed bremsstrahlung spectra, without including the quark effects,
namely, using the free nucleon magnetic moments,
$\mu_p = 2.792 84734462$ and $\mu_n = -1.91304273$.
It turns out that we need to multiply a constant, $c =  1.296296296$, to reproduce the cross
sections data.
Then, we also have been able to reconstruct the old results
in Ref.~\cite{Maydanyuk.2023.PRC.delta} without the $\Delta$ resonances.
This spectrum calculated without the quark effects 
is in close agreement with the experimental data
[see Fig.~\ref{fig.analysis.197Au.1}~(a)].
The same constant, $c =  1.296296296$, will be multiplied for all the calculations in the following.

Then, we calculate the spectrum with the quark effects.
For this purpose, we use the in-medium modified proton and neutron magnetic moments.
We follow the procedure in Ref.~\cite{Tsushima:2020gun}.
The in-medium nucleon magnetic moments are estimated 
using the averaged nucleon density for the nucleus.
For \isotope[197]{Au} the average nucleon density is calculated by the QMC model as
$\rho = 0.78286\, \rho_{0}$ with $\rho_{0} = 0.15\; \mbox{\rm fm}^{-3}$.
We find the in-medium magnetic moments of proton and neutron with the
average nuclear density of $^{197}$Au 
respectively as
$\mu_{\rm p}^{*} / \mu_{\rm p}^{(0)} = 
  1.078\: 154$ and
$\mu_{\rm n}^{*} / \mu_{\rm n}^{(0)} = 
1.077\; 861$.

The calculated bremsstrahlung cross section for the nucleus \isotope[197]{Au}
at the proton beam
energy $E_{\rm p}=190$~MeV is presented in Fig.~\ref{fig.analysis.197Au.1}~(a)
(blue dashed line),
with the in-medium modified nucleon magnetic moments.
Also we add the results without the quark effects
(red solid line).

One can see that the difference between the spectra for
the model with and without the quark effects 
is not so large.
However, the amount of the difference is stable in the full energy region of photons.
This result confirms a possibility to observe the quark effects in nuclei
in the spectra of bremsstrahlung in proton-nucleus scattering. 
Based on these results, we address the following question, namely,
for which nuclei the quark effects should be better visible.
To answer this question is one of the main aims of
the present study.

\vspace{1.5mm}
\emph{Strategy for finding quark effects.
Quark effects in middle mass nuclei and spectra for calcium \isotope[40]{Ca}.
\label{sec.analysis.40Ca}}\;
We find that, with the in-medium nucleon magnetic moments,
the incoherent contribution is indeed modified, while the coherent contribution is nearly unmodified.
According to Ref.~\cite{Maydanyuk.2023.PRC.delta},
we know that the incoherent contribution is much larger than
the coherent one as mentioned already.
For example, for \isotope[197]{Au}
the ratio,  (incoherent bremsstrahlung)/(coherent bremsstrahlung)
is about $10^{+6}$--$10^{+7}$~\cite{Maydanyuk.2023.PRC.delta}.
Thus, the role of the coherent terms are negligible in the full spectrum.
If we normalize the calculated spectrum with and without the quark effects
for the same experimental data point for \isotope[197]{Au},
these spectra (shapes) become similar.
We can conclude that it is better to find another nucleus,
where magnitudes of the incoherent and coherent contributions are similar.
It turns out that the incoherent contribution decreases in comparison
with the coherent one, when we study the nuclei from heavy nuclei
to light~\cite{Maydanyuk.2023.PRC.delta}.
Thus, we search for middle mass nuclei, and try the \isotope[40]{Ca} nucleus.

The calculation of the in-medium nucleon magnetic moments for this nucleus
is similar as described above.
The in-medium nucleon magnetic moments are obtained using
the average nuclear density of \isotope[40]{Ca},
$\rho = 0.71148\, \rho_{0}$. 
We obtain the in-medium proton and neutron magnetic moments respectively as
$\mu_{\rm p}^{*} / \mu_{\rm p}^{(0)} = 0.1071$ and
$\mu_{\rm n}^{*} / \mu_{\rm n}^{(0)} = 1.066$.
Calculations of the bremsstrahlung spectra for  \isotope[40]{Ca}
at the proton beam energy of $E_{\rm p}=190$~MeV with and without the quark effects
are presented in Fig.~\ref{fig.analysis.197Au.1}~(b).
However, the magnitudes of the incoherent and coherent bremsstrahlung are similar, 
at the low photon energy region
--- up to 0.4~MeV (cannot be seen in the figure), although at higher energies the incoherent
contribution is dominant.
Thus, \isotope[40]{Ca} is not good enough to observe the quark effects in the spectra.

\vspace{1.5mm}
\emph{Quark effects in light nuclei. Spectra for oxygen \isotope[16]{O}.
\label{sec.analysis.16O}}\;
As the next step, we study smaller mass nuclei, and choose
oxygen \isotope[16]{O} for our continuous analysis.
For \isotope[16]{O} we expected to see an increase of the coherent contribution
rate that may be able to give larger deformation of the full spectrum.
But our estimates shown in Fig.~\ref{fig.analysis.197Au.1}~(c),
indicate that the visibility of the quark effects 
is only slightly better than that for \isotope[40]{Ca},
because the coherent and incoherent contributions are similar
only at photon energies less than 5~MeV.
Thus, we further continue to search for a better, lighter nucleus.
The calculated in-medium magnetic moments of proton and
neutron for the \isotope[16]{O} nucleus with the average nuclear density of
$\rho = 0.61242 \, \rho_{0}$, are respectively,
$\mu_{\rm p}^{*} / \mu_{\rm p}^{(0)} = 
1.06\, 286$ and
$\mu_{\rm n}^{*} / \mu_{\rm n}^{(0)} = 
1.05\, 750$.

\vspace{1.5mm}
\emph{Quark effects in light nuclei.
Spectra for isotopes of carbon \isotope[12]{C} and \isotope[18]{C}.
\label{sec.analysis.12C}}\;
Our estimates of the spectra for proton scattering on deuterons show the dominant
role of the coherent contribution of bremsstrahlung.
However, we need to study the nucleus where the coherent and
incoherent contributions
have similar magnitudes, but not at the very small photon energies.
Therefore, we expect that the nucleus mass range between the deuteron and oxygen
should be suitable.
It turns out that isotopes of carbon are suitable for those conditions.
In addition, the nuclear density of deuteron is too small
to be able to apply properly the QMC model (mean field model).

We calculate the in-medium magnetic moments of protons and neutrons in the nucleus
\isotope[12]{C}.
The in-medium nucleon magnetic moments are obtained 
by using the average nuclear density of \isotope[12]{C}
with $\rho = 0.53277\, \rho_{0}$.
We obtain the in-medium magnetic moments of
protons and neutrons 
respectively as
$\mu_{\rm p}^{*} / \mu_{\rm p}^{(0)} = 
1.055\, 091$ and
$\mu_{\rm n}^{*} / \mu_{\rm n}^{(0)} = 
1.050\, 417$.
While, for the \isotope[18]{C} nucleus, the in-medium proton and neutron magnetic moments are calculated with
the average nucleon density of $\rho=0.52137 \, \rho_0$, and we get respectively,
$\mu_p^*/\mu_p = 1.054\, 102$ 
and $\mu_n^*/\mu_n = 1.049\, 399$. 

\begin{figure}[t]
\centerline{%
\includegraphics[width=65mm]{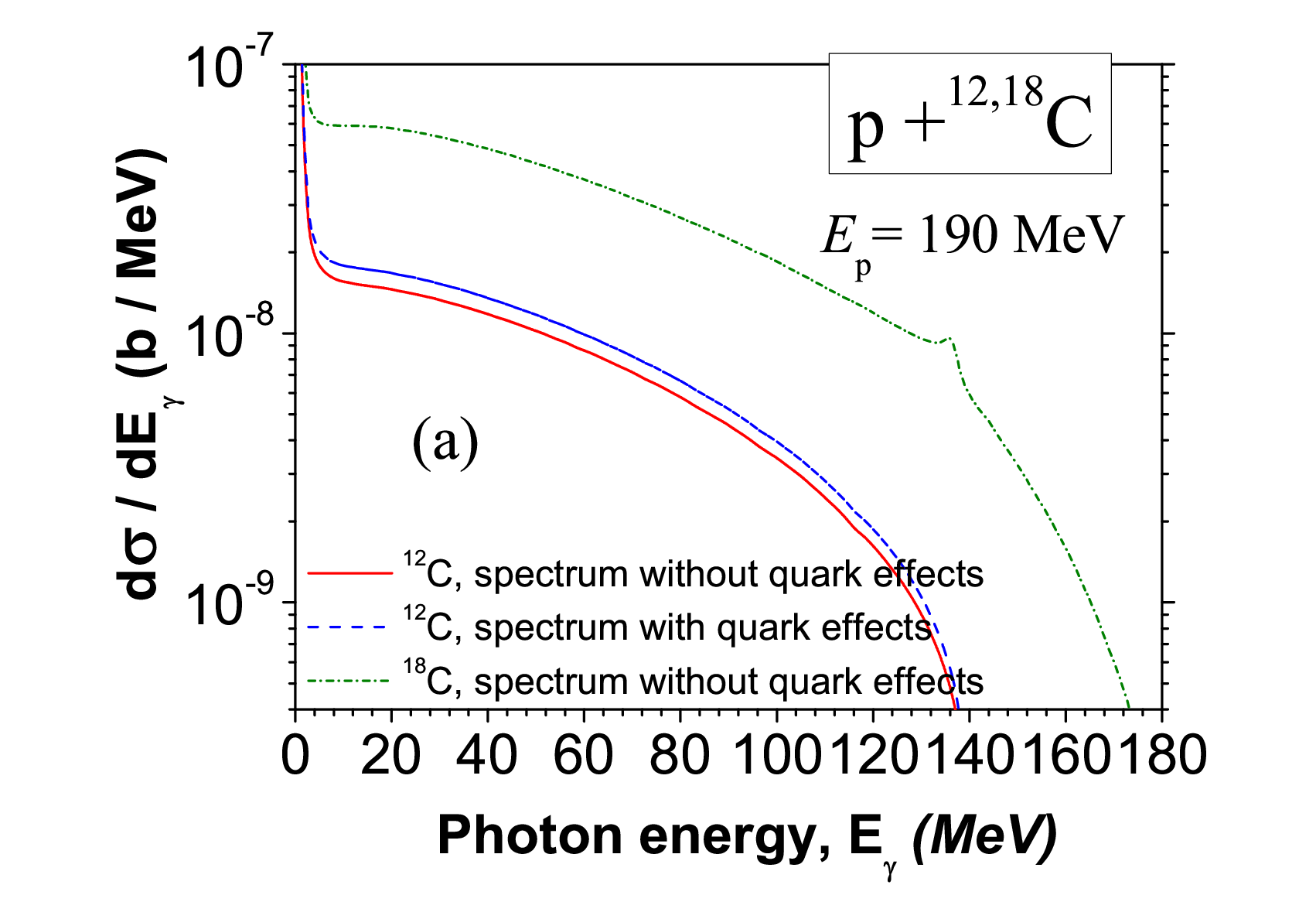}
\hspace{-7.0mm}\includegraphics[width=65mm]{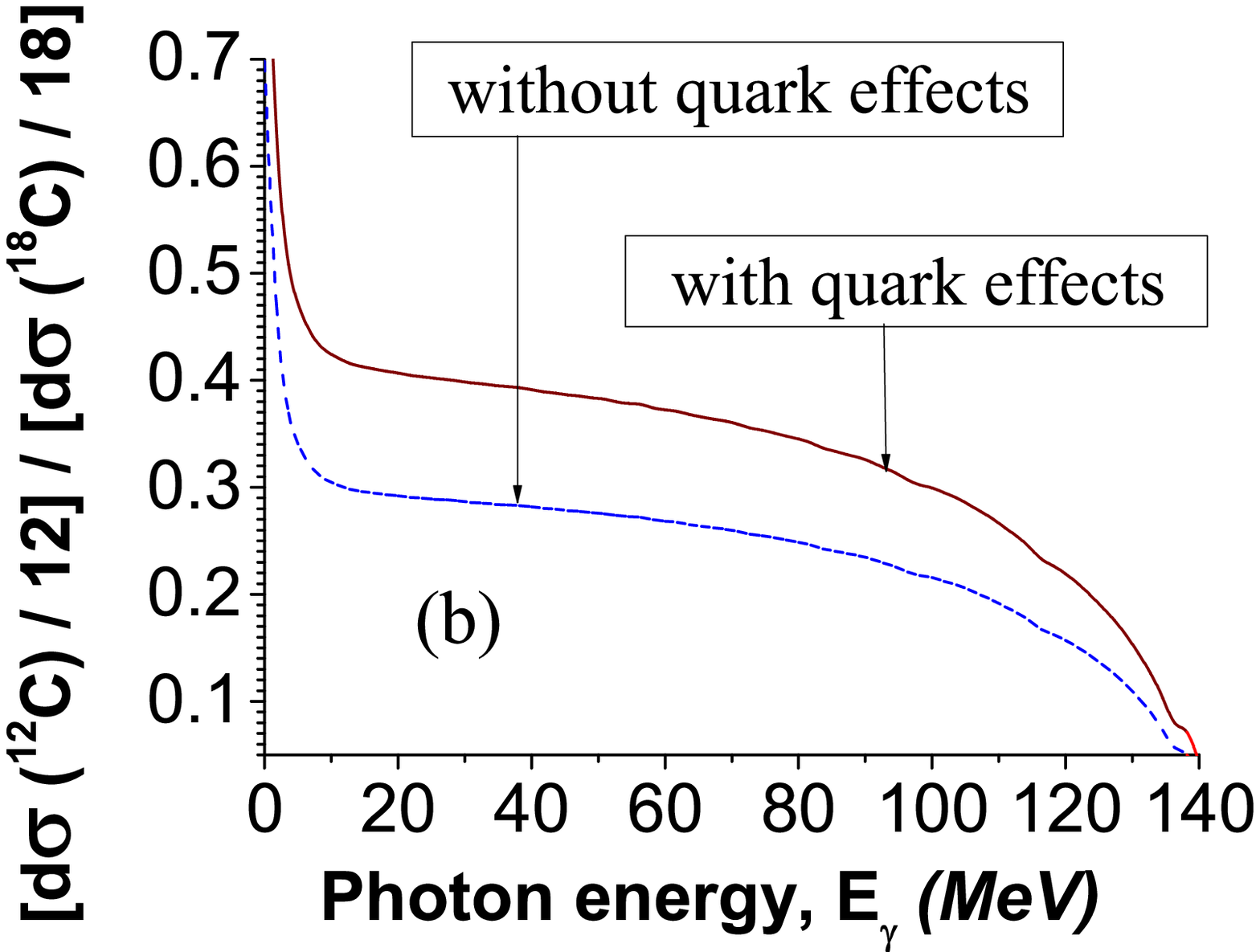}
\hspace{-3.0mm}\includegraphics[width=65mm]{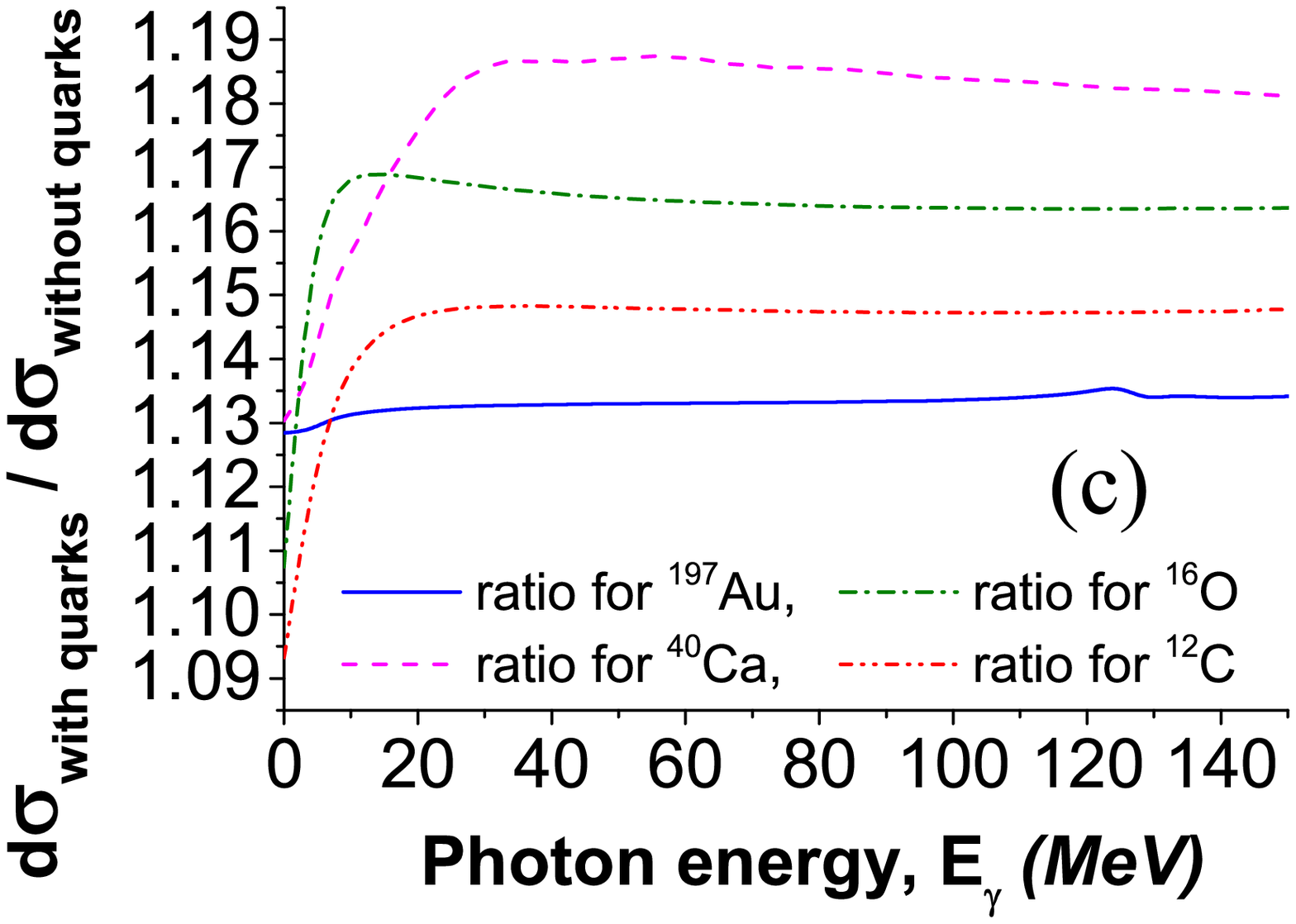}}
\vspace{-4mm}
\caption{\small (Color online)
Panel (a):
Calculated bremsstrahlung differential cross sections (with the coherent plus incoherent terms)
in the scattering of protons off the \isotope[12,18]{C} nuclei at the proton beam energy of
$E_{\rm p}=190$~MeV.
Panel (b): Ratios for the differential cross sections normalized for one nucleon,
[$d\sigma(^{12}$C)/12]/[$d\sigma(^{18}$C)/18]
with and without the quark effects,
where the \isotope[18]{C} cross section has minimum quark effects (nearly absent), 
and the ratio is enhanced when we include the quark effects for the both cases
for \isotope[12]{C} and \isotope[18]{C}, due to the quark effects for \isotope[12]{C}.
Panel (c): Ratios for the bremsstrahlung differential cross sections
[with the quark effects]/[without the quark effects]
for different nuclei. 
\label{fig.analysis.12C.1}}
\end{figure}

With the in-medium nucleon magnetic moments calculated as above,
the calculated differential cross sections
$d\sigma$(\isotope[12,18]{C})/$dE_\gamma$
are shown in Fig.~\ref{fig.analysis.12C.1}~(a),
at the proton beam energy $E_{\rm p}=190$~MeV
with and without the quark effects.
This is because, after calculations for several carbon isotopes, we
find that the isotope \isotope[18]{C} nucleus has the minimal
incoherent bremsstrahlung contribution,
and thus the role of the quark effects should be minimal for the incoherent bremsstrahlung
contribution in comparison with the other isotopes of carbon.
Then, the normalized cross section ratio for one nucleon,
[$d\sigma (^{12}$C)/12]/[$d\sigma(^{18}$C)/18]
is expected to be enhanced, since the $^{12}$C spectrum contains larger contribution
from the quark effects than that of $^{18}$C. 
Indeed, this effect can be seen in Fig.~\ref{fig.analysis.12C.1}~(b).
%

\vspace{.05cm}

The result shown in Fig.~\ref{fig.analysis.12C.1}~(b) is independent of the
multiplied constant $c = 1.296296296$ in the calculations.
Thus, we expect that they can provide sufficiently solid information
about the role of the quark effects, and this fact can be proposed for
future experimental confirmation.

One can see in Fig.~\ref{fig.analysis.12C.1}~(b) that for both with and without the quark effects,
the full spectrum for \isotope[18]{C} is larger than for \isotope[12]{C}.
To explain this result, we compare the ratio between the coherent contributions for \isotope[12]{C}
and \isotope[18]{C}.
The coherent contribution is proportional to effective electric charge $Z_{\rm eff}$ (see Eq.~(S29)).
It turns out that effective electric charge for \isotope[18]{C}
[$Z_{\rm eff} (\isotope[18]{C}) = 0.63 157 894$] is larger
than for \isotope[12]{C} [$Z_{\rm eff} (\isotope[12]{C}) = 0.46 153 846$]. 
We obtain the ratio $Z_{\rm eff} (\isotope[12]{C}) / Z_{\rm eff} (\isotope[18]{C}) = 0.73 076 922$.
Thus, one can naively expect that the normalized \isotope[18]{C} spectra
is larger than those of \isotope[12]{C}.
However, the result shown in Fig.~\ref{fig.analysis.12C.1}~(b) is obtained
with more accurate calculation including other effects, such as the photon energy
$E_\gamma$ dependence for the incoherent and coherent terms.
%
In Fig.~\ref{fig.analysis.12C.1}~(c) we show the ratios of differential cross sections,
[with the quark effects]/[without the quark effects],
for nuclei $^{12}$C, $^{16}$O, $^{40}$Ca, and $^{197}$Au,
where these cases are enhanced and larger than the unity, due to the enhancement of the
nucleon magnetic moments in each nucleus.

\vspace{1.5mm}
\emph{Conclusions.
\label{sec.conclusions}}\;
To conclude, we have reported on the new possible physical observation
of quark effects in nuclei in the bremsstrahlung photon emission in nuclear reactions.
We have extended the existing established bremsstrahlung model
for the proton-nucleus scattering~\cite{Maydanyuk.2023.PRC.delta},
by incorporating the in-medium modified nucleon magnetic moments
in nuclei based on the quark-meson coupling model~\cite{Tsushima:2020gun}.
To explore the quark effects in the calculated bremsstrahlung cross sections
for different nuclei from light to heavy ones,
we have established a possibility of observing the quark effects for the first time.
Importantly, this can be measured experimentally.
Moreover, we emphasize that, such quark effects visibly modify the incoherent photon
emission, while only slightly for the coherent emission.
Therefore, the shape of the full bremsstrahlung cross section is modified
maximally after the inclusion of the quark effects
(in-medium modified nucleon magnetic moments in nuclei) in the
model, if one uses nuclei with minimal difference in the magnitudes of the
coherent and incoherent bremsstrahlung contributions.
In this study, we have found that the carbon isotopes satisfy this condition.
This is the first time finding, and also our proposal
to explore the quark effects in nuclei via the nuclear bremsstrahlung photon emission
both in theoretically and experimentally,
although bremsstrahlung emission in nuclear physics at low energies
have been studied for more than 90 years.

Analyzing nuclei from light to heavy masses,
we find that the isotope \isotope[18]{C} has the minimal incoherent bremsstrahlung
contribution, and thus the role of the quark effects should be minimal
in comparison with the other isotopes of carbon,
where in the latter, the role of incoherent bremsstrahlung is larger.
In particular, the normalized ratios for one nucleon between the spectra for \isotope[18]{C} and
those for \isotope[12]{C} with and without the quark effects,
may provide sufficiently solid information on the role of quark effects in nuclei
as shown in Fig.~\ref{fig.analysis.12C.1}~(b).
This is a new and novel way that we propose to use for future experimental measurements
in search for the possible quark effects in nuclear reactions.


\section*{Acknowledgements
\label{sec.acknowledgements}}

S.P.M. thanks Sun Yat-Sen University for warm hospitality and support.
K.T.~was supported by Conselho Nacional de Desenvolvimento
Cient\'{i}fico e Tecnol\'ogico (CNPq, Brazil), Processes No.~313063/2018-4,
No.~426150/2018-0, No.~304199/2022-2,
and FAPESP Process No.~2019/00763-0 and No.~2023/07313-6.
G.R.~was supported by the National Research Foundation of Korea, Grant  No.~RS-2021-NR060129.
The work of authors was in the projects of
Instituto Nacional de Ci\^{e}ncia e
Tecnologia - Nuclear Physics and Applications
(INCT-FNA), Brazil, Process No.~464898/2014-5.




%
%
%
%

\widetext
\clearpage 
~\vspace{1cm} 
\begin{center}
%
%
  \textbf{\large
    Manifestation of quark effects in nuclei via bremsstrahlung analysis in the proton-nucleus scattering \\
    \vspace{.2cm}
    Supplemental Material}
\end{center}
\setcounter{equation}{0}
\setcounter{figure}{0}
\setcounter{table}{0}
\setcounter{page}{1}
\makeatletter
\renewcommand{\theequation}{S\arabic{equation}}
\renewcommand{\thefigure}{S\arabic{figure}}
\renewcommand{\bibnumfmt}[1]{[S#1]}
\renewcommand{\citenumfont}[1]{S#1} 




\section{Model
\label{sec.model}}

\subsection{Hamiltonian of proton-nucleus scattering and
bremsstrahlung photon emission operators
\label{sec.2.1}}

Let us consider scattering of proton on a nucleus {with the mass number $A$
in the laboratory frame.
We write the Hamiltonian including photon emission as
many-nucleon generalization of Pauli equation
(see Refs.~\cite{Maydanyuk.2012.PRC,Maydanyuk_Zhang.2015.PRC,Maydanyuk_Zhang_Zou.2016.PRC},
and references therein):
\begin{equation}
\begin{array}{lcl}
\vspace{1mm}
  \hat{H} & = &
  \biggl\{
    \displaystyle\frac{1}{2m_{\rm p}} \Bigl( \vu{p}_{\rm p} - \displaystyle\frac{z_{\rm p}e}{c} \vb{A}_{\rm p} \Bigr)^{2} +
    z_{\rm p}e\, A_{\rm p,0} - \displaystyle\frac{z_{\rm p}e \hbar}{2m_{\rm p} c}\; \bigl( \hat{\sigmabf}_{\rm p} \cdot \vb{rot\, A}_{\rm p} \bigr)
  \biggr\}\; 
  \\
  & + &
  \displaystyle\sum_{i=1}^{A}
  \biggl\{
    \displaystyle\frac{1}{2m_{i}} \Bigl( \vu{p}_{i} - \displaystyle\frac{z_{i}e}{c} \vb{A}_{i} \Bigr)^{2} +
    z_{i}e\, A_{i,0} - \displaystyle\frac{z_{i}e \hbar}{2m_{i}c}\; \bigl( \hat{\sigmabf}_{i} \cdot \vb{rot\, A}_{i} \bigr)
  \biggr\} +
  V\, (\vb{r}_{1} \ldots \vb{r}_{A}, \vb{r}_{\rm p}).
\end{array}
\label{eq.pauli.2}
\end{equation}
Here, $e$ is the positron charge,
$m_{i}$ and $z_{i}$ are the mass and electric charge of $i$-th nucleon in the nucleus,
$\vu{p}_{i} = -i\hbar\, \vb{d}/\vb{dr}_{i} $ is the $i$-th nucleon momentum operator,
$V(\vb{r}_{1} \ldots \vb{r}_{A}, \vb{r}_{\rm p})$ is a generic form of the potential
for the interactions among nucleons in the nucleus and the proton in the beam,
$\hat{\sigmabf}_{\rm p}$ and $\hat{\sigmabf}_{i}$ are
the spin operators acting on proton in the beam and
$i$-th nucleon in the nucleus, respectively
(see Eqs.~(12)--(20) in Ref.~\cite{Maydanyuk.2012.PRC}, for details), 
$A_{i} \equiv (\vb{A}_{i}, A_{i,0})$ is a electromagnetic potential
generated by the by the $i$-th nucleon, and
``$A$'' appearing in the upper limit of the summation
is the mass number of the target nucleus.

The Hamiltonian~(\ref{eq.pauli.2}) can be rewritten as
\begin{equation}
\begin{array}{lcl}
  \hat{H} = \hat{H}_{0} + \hat{H}_{\gamma},
\end{array}
\label{eq.pauli.4}
\end{equation}
where
\begin{equation}
\begin{array}{lllll}
\vspace{1mm}
  \hat{H}_{0} & = &
    \displaystyle\frac{1}{2m_{\rm p}}\, \vu{p}_{\rm p}^{2} +
  \displaystyle\sum_{i=1}^{A} \displaystyle\frac{1}{2m_{i}}\, \vu{p}_{i}^{2} +
  V (\vb{r}_{1} \ldots \vb{r}_{A}, \vb{r}_{\rm p}), \\
\vspace{1mm}
  \hat{H}_{\gamma} & = &
  \biggl\{
    - \displaystyle\frac{z_{\rm p} e}{m_{\rm p}c}\; \bigl( \vu{p}_{\rm p} \cdot \vb{A}_{\rm p} \bigr) +
    \displaystyle\frac{z_{\rm p}^{2}e^{2}}{2m_{\rm p}c^{2}} \vb{A}_{\rm p}^{2} -
    \displaystyle\frac{z_{\rm p}e\hbar}{2m_{\rm p}c}\, \bigl( \hat{\sigmabf}_{\rm p} \cdot \vb{rot A}_{\rm p} \bigr) +
    z_{\rm p}e\, A_{\rm p,0}
  \biggr\}\; 
  \\
  & + &
  \displaystyle\sum_{i=1}^{A}
  \biggl\{
    - \displaystyle\frac{z_{i} e}{m_{i}c}\; \bigl( \vu{p}_{i} \cdot \vb{A}_{i} \bigr) +
    \displaystyle\frac{z_{i}^{2}e^{2}}{2m_{i}c^{2}} \vb{A}_{i}^{2} -
    \displaystyle\frac{z_{i}e\hbar}{2m_{i}c}\, \bigl( \hat{\sigmabf}_{i} \cdot \vb{rot A}_{i} \bigr) +
    z_{i}e\, A_{i,0}
  \biggr\},
\end{array}
\label{eq.pauli.5}
\end{equation}
where, $\hat{H}_{0}$ is the Hamiltonian  of nucleons in  the target nucleus
without the photon emission, while
$\hat{H}_{\gamma}$ is the Hamiltonian associated with the bremsstrahlung photon emission.

We use the notation for the magnetic moments:
$\mu_{i}^{\rm (Dir)} = z_{i}\, e\hbar / (2m_{i}c)$ for $i$-th nucleon as
$\mu_{i}^{\rm (Dir)} \to \mu_{i}\, \mu_{0}$,
where
$\mu_{0} = e\hbar / (2m_{\rm p}c)$ is the nuclear magneton,
$\mu_{\rm p} = 2.792\; 847\; 344\; 62$ and $\mu_{\rm n} = -1.913\; 042\; 73$
are the total magnetic moments of proton and neutron, respectively, and
thus $\mu_{i}$ is the magnetic moment of $i$-th nucleon in the nucleus
in units of $\mu_{0}$.
We neglect the terms proportional to $\vb{A}_{j}^{2}$ and $A_{j,0}$,
and use the Coulomb gauge.%

Operator of photon emission~(\ref{eq.pauli.5}) is transformed as
\begin{equation}
\begin{array}{llll}
  \hat{H}_{\gamma} & = &
  \biggl\{
    - \displaystyle\frac{z_{\rm p} e}{m_{\rm p}c}\; \bigl( \vu{p}_{\rm p} \cdot \vb{A}_{\rm p} \bigr) -
    \mu_{0}\, \mu_{\rm p}\, \bigl( \hat{\sigmabf}_{\rm p} \cdot \vu{H}_{\rm p} \bigr)
  \biggr\} +
  \displaystyle\sum_{i=1}^{A}
  \biggl\{
    - \displaystyle\frac{z_{i} e}{m_{i}c}\; \bigl( \vu{p}_{i} \cdot \vb{A}_{i} \bigr) -
    \mu_{0}\, \mu_{i}\, \bigl( \hat{\sigmabf}_{i} \cdot \vu{H}_{i} \bigr)
  \biggr\},
\end{array}
\label{eq.2.2.6}
\end{equation}
where%
\footnote{One can see that the formulation  violates gauge invariance.
Moreover, in the shell model of the nucleus, the full magnetic moments
of protons and neutrons in the nucleus are depended on the shells in the nucleus.
This formulation also violates gauge invariance.
However the level of accuracy for determining some spectroscopic properties
of nuclei in the nuclear shell model framework is sufficiently good.}
\begin{equation}
  \vu{H} = \vb{rot\: A} = \curl{\vb{A}} \bigr.
\label{eq.2.2.5}
\end{equation}

\subsection{Formalism in coordinate space representation
\label{sec.2.3}}

Using the following substitution for the electromagnetic potential,
\begin{equation}
\begin{array}{lcl}
  \vb{A} & = &
  \displaystyle\sum\limits_{\alpha=1,2}
    \sqrt{\displaystyle\frac{2\pi\hbar c^{2}}{w_{\rm ph}}}\; \vb{e}^{(\alpha)\,*}
    e^{-i\, \vb{k_{\rm ph}r}},
\end{array}
\label{eq.2.3.1}
\end{equation}
we obtain
\begin{equation}
\begin{array}{lcl}
  \vu{H} & = &
  \curl{\vb{A}} =
  \sqrt{\displaystyle\frac{2\pi\hbar c^{2}}{w_{\rm ph}}}\,
    \displaystyle\sum\limits_{\alpha=1,2}
    \Bigl\{ -i\, e^{-i\, \vb{k_{\rm ph}r}}\, \bigl[ \vb{k_{\rm ph}} \times \vb{e}^{(\alpha)\,*}
\bigr] +
      e^{-i\, \vb{k_{\rm ph}r}}\, \bigl[ \curl{\vb{e}^{(\alpha)\,*}} \bigr] \Bigr\}.
\end{array}
\label{eq.2.3.2}
\end{equation}
Here,
$\vb{e}^{(1)}$ and $\vb{e}^{(2)}$ are the two independent polarization vectors
of photon
[$\vb{e}^{(\alpha)*} = \vb{e}^{(\alpha)}$, $\alpha=1,2$],
$\vb{k}_{\rm ph}$ is the wave vector of the photon and
$w_{\rm ph} = k_{\rm ph} c = \bigl| \vb{k}_{\rm ph}\bigr|c$.
The vectors $\vb{e}^{(\alpha)}$ are perpendicular to $\vb{k}_{\rm ph}$ in the Coulomb gauge,
and satisfy Eq.~(8)
in Ref.~\cite{Liu_Maydanyuk_Zhang_Liu.2019.PRC.hypernuclei}.
In the following, we shall keep explicitly
$\hbar$ and $c$.
Note that the following relations (see Eqs.~(7) in Ref.~\cite{Maydanyuk.2012.PRC}):
\begin{equation}
\begin{array}{lclc}
   \vb{k}_{\rm ph} \times \vb{e}^{(1)}  = k_{\rm ph}\, \vb{e}^{(2)},  \hspace{.15cm}&
   \vb{k}_{\rm ph} \times \vb{e}^{(2)}  = -\, k_{\rm ph}\, \vb{e}^{(1)}, \hspace{.15cm} &
   \vb{k}_{\rm ph} \times \vb{e}^{(3)}  = 0,  \hspace{.15cm}&
  \displaystyle\sum\limits_{\alpha=1,2} \Bigl[ \vb{k}_{\rm ph} \times \vb{e}^{(\alpha)} \Bigr] = k_{\rm ph}\, (\vb{e}^{(2)} - \vb{e}^{(1)}).
\end{array}
\label{eq.2.3.3}
\end{equation} 
Substituting formulas (\ref{eq.2.3.1}) and (\ref{eq.2.3.2})
to Eq.~(\ref{eq.2.2.6}) for the photon emission operator, we get
\begin{equation}
\begin{array}{lcl}
  \hat{H}_{\gamma} & = &
  \sqrt{\displaystyle\frac{2\pi\hbar c^{2}}{w_{\rm ph}}}\; \mu_{0}\,
  \displaystyle\sum\limits_{\alpha=1,2}
    e^{-i\, \vb{k_{\rm ph}r}_{i}}\,
  \biggl\{
    i\, 2 z_{\rm p}\: \bigl( \vb{e}^{(\alpha)} \cdot \grad_{i} \bigr) +
    \mu_{\rm p}\, 
    \Bigl[ i\, \bigl( \hat{\sigmabf}_{\rm p} \cdot \bigl[ \vb{k_{\rm ph}} \times \vb{e}^{(\alpha)} \bigr] \bigr) - 
      \bigl( \hat{\sigmabf}_{\rm p} \cdot \bigl[ \grad_{\rm p} \times \vb{e}^{(\alpha)} \bigr] \bigr) 
      \Bigr]
  \biggr\}\; 
  \\
  & + &
  \sqrt{\displaystyle\frac{2\pi\hbar c^{2}}{w_{\rm ph}}}\; \mu_{0}\,
  \displaystyle\sum_{i=1}^{A}
  \displaystyle\sum\limits_{\alpha=1,2}
    e^{-i\, \vb{k_{\rm ph}r}_{i}}\;
    \biggl\{
      i\, \displaystyle\frac{2 z_{i} m_{\rm p}}{m_{Ai}}\: \bigl( \vb{e}^{(\alpha)} \cdot \grad_{i} \bigr) +
      \mu_{i}\, 
      \Bigl[ i\, \bigl( \hat{\sigmabf}_{i} \cdot \bigl[ \vb{k_{\rm ph}} \times \vb{e}^{(\alpha)} \bigr] \bigr) - 
      \bigl( \hat{\sigmabf}_{i} \cdot \bigl[ \grad_{i} \times \vb{e}^{(\alpha)} \bigr] \bigr) \Bigr]
    \biggr\}.
\end{array}
\label{eq.2.3.4}
\end{equation}

\subsection{Operator of photon emission with the relative coordinates
\label{sec.2.5}}

We define 
the coordinates as follows;
$\vb{R}_{A}$ the target nucleus (A) center-of-mass (CM) coordinate in laboratory frame,
$\vb{R}$ the total (beam proton + target nucleus) system CM coordinate in laboratory frame,
$\vb{r}$ the relative coordinate between the beam proton and the target nucleus
CM coordinate ($\vb{R}_{A}$),
and $\rhobf_{A i}$ the $i$-th nucleon relative coordinate from the CM coordinate of the target nucleus ($\vb{R}_{A}$).
Following Ref.~\cite{Maydanyuk.2012.PRC} (see Eqs.~(3), (4) in the
paper, and also see Appendix~A in Ref.~\cite{Liu_Maydanyuk_Zhang_Liu.2019.PRC.hypernuclei}),
we find, $\vb{r} = \vb{r}_{\rm p} - \vb{R}_{A}$,
and $\rhobf_{Ai} = \vb{r}_{Ai} - \vb{R}_{A}$ for nucleons in the nucleus $A$.
We calculate the momenta $\vu{P}$, $\vu{p}$, $\vb{\tilde{p}}_{Ai}$
corresponding to $\vb{R}$, $\vb{r}$,
$\rhobf_{A i}$ for $i = 1 \ldots A-1$
($\vu{p}_{i} \equiv -i\hbar\, \vb{d}/\vb{dr}_{i}$), and rewrite formalism
above.\footnote{One can write useful formulas for the nucleons
in the the target nucleus with the nucleon number $A$ as
\begin{equation}
\begin{array}{lllll}
  \rhobf_{AA} = -\, \displaystyle\frac{1}{m_{AA}} \displaystyle\sum_{k=1}^{A-1} m_{Ak}\,
  \rhobf_{Ak}, &
\hspace{3ex}
  \vu{p}_{AA} =
    \displaystyle\frac{m_{AA}}{m_{A} + m_{p}}\, \vu{P} -
    \displaystyle\frac{m_{AA}}{m_{A}}\,\vu{p} -
    \displaystyle\frac{m_{AA}}{m_{A}}\, \displaystyle\sum_{k=1}^{A-1} \vb{\tilde{p}}_{Ak}.
\end{array}
\label{eq.footnote.1}
\end{equation}
}
%

Following Ref.~\cite{Maydanyuk.2023.PRC.delta},
we calculate operator of the bremsstrahlung photon emission
in the scattering of proton on the nucleus $A$
in the laboratory frame as
\begin{equation}
  \hat{H}_{\gamma} =
  \hat{H}_{P} + \hat{H}_{p} + \Delta \hat{H}_{\gamma E} + \Delta \hat{H}_{\gamma M} + \hat{H}_{k}.
\label{eq.app.OpEmission.pA.1}
\end{equation}
For the proton-nucleus scattering, such operators were calculated in
Refs.~\cite{Maydanyuk.2023.PRC.delta,Liu_Maydanyuk_Zhang_Liu.2019.PRC.hypernuclei}.
We will write down the explicit expressions in detail.
The term related with the motion of the full proton-nucleus system is given by,
\begin{equation}
\begin{array}{lcl}
\vspace{-0.1mm}
  \hat{H}_{P} & = &
  -\, \sqrt{\displaystyle\frac{2\pi c^{2}}{\hbar w_{\rm ph}}}\;
  \mu_{0}\, \displaystyle\frac{2 m_{\rm p}}{m_{\rm p} + m_{A}}\;
  e^{-i\, \vb{k_{\rm ph}} \vb{R}}
  \displaystyle\sum\limits_{\alpha=1,2}
  \biggl\{
    z_{\rm p}\, e^{-i\, c_{A}\, \vb{k_{\rm ph}} \vb{r}} +
    e^{i\, c_{\rm p}\, \vb{k_{\rm ph}} \vb{r} }  \displaystyle\sum_{i=1}^{A} z_{i}\, e^{-i\, \vb{k_{\rm ph}} \rhobf_{Ai}}
  \biggr\}\, \bigl( \vb{e}^{(\alpha)} \cdot \vu{P} \bigr)\; 
  \\
  & - &
  \sqrt{\displaystyle\frac{2\pi c^{2}}{\hbar w_{\rm ph}}}\;
    \displaystyle\frac{i\, \mu_{0}}{m_{\rm p} + m_{A}}\;
    e^{-i\, \vb{k_{\rm ph}} \vb{R}}\,
    \displaystyle\sum\limits_{\alpha=1,2}
    \biggl(
    \Bigl\{
      e^{-i\, c_{A}\, \vb{k_{\rm ph}} \vb{r}}\, \mu_{\rm p}\, m_{\rm p}\, \hat{\sigmabf}_{\rm p}  +
      e^{i\, c_{\rm p}\, \vb{k_{\rm ph}} \vb{r}}\,
      \displaystyle\sum_{i=1}^{A}
      \mu_{\rm Ai}\, m_{\rm Ai}\, e^{-i\, \vb{k_{\rm ph}} \rhobf_{\rm Ai}}\, \hat{\sigmabf}_{i}
  \Bigr\}\, \cdot \bigl[ \vu{P} \times \vb{e}^{(\alpha)} \bigr] \biggr),
\end{array}
\label{eq.app.OpEmission.pA.2}
\end{equation}
The coherent term is given by
(see Eq.~(18) for scattering of $\alpha$ particles on nuclei
in Ref.~\cite{Liu_Maydanyuk_Zhang_Liu.2019.PRC.hypernuclei})
\begin{equation}
\begin{array}{lll}
\vspace{-0.1mm}
  & \hat{H}_{p} =
  -\, \sqrt{\displaystyle\frac{2\pi c^{2}}{\hbar w_{\rm ph}}}\;
  2\, \mu_{0}\,  m_{\rm p}\,
  e^{-i\, \vb{k_{\rm ph}} \vb{R}}
  \displaystyle\sum\limits_{\alpha=1,2}
  \biggl\{
    \displaystyle\frac{z_{\rm p}}{m_{\rm p}}\, e^{-i\, c_{A} \vb{k_{\rm ph}} \vb{r}} -
    e^{i\, c_{\rm p} \vb{k_{\rm ph}} \vb{r}}\,  \displaystyle\frac{1}{m_{A}}\,
      \displaystyle\sum_{i=1}^{A} z_{Ai}\, e^{-i\, \vb{k_{\rm ph}} \rhobf_{Ai}}
  \biggr\}\; \bigl( \vb{e}^{(\alpha)} \cdot \vu{p} \bigr)\; 
  \\
  - &
  i\, \mu_{0}\, \sqrt{\displaystyle\frac{2\pi c^{2}}{\hbar w_{\rm ph}}}\;
  e^{-i\, \vb{k_{\rm ph}} \vb{R}}
  \displaystyle\sum\limits_{\alpha=1,2}
  \biggl(
  \Bigl\{
    e^{-i\, c_{A} \vb{k_{\rm ph}} \vb{r}}\, \mu_{\rm p}\, \hat{\sigmabf}_{\rm p} -
    e^{i\, c_{\rm p} \vb{k_{\rm ph}} \vb{r}} \displaystyle\frac{1}{m_{A}}
      \displaystyle\sum_{i=1}^{A} \mu_{Ai}\, m_{Ai}\; e^{-i\, \vb{k_{\rm ph}} \rhobf_{Ai}}\, \hat{\sigmabf}_{i} \biggr\}
  \cdot \bigl[ \vu{p} \times \vb{e}^{(\alpha)} \bigr]
  \biggr).
\end{array}
\label{eq.app.OpEmission.pA.3}
\end{equation}
The incoherent terms of electric and magnetic types are written by
(see Eqs.~(12), (13) for proton-nucleus scattering in Ref.~\cite{Maydanyuk.2023.PRC.delta}),
\begin{equation}
\begin{array}{lcl}
\vspace{0.5mm}
  \Delta \hat{H}_{\gamma E} & = &
  -\, \sqrt{\displaystyle\frac{2\pi c^{2}}{\hbar w_{\rm ph}}}\:
    2\, \mu_{0}\, e^{-i\, \vb{k_{\rm ph}}\vb{R}}\,
  \displaystyle\sum\limits_{\alpha=1,2}
  \biggl\{
    e^{i\, c_{\rm p}\, \vb{k_{\rm ph}} \vb{r}}\,
    \displaystyle\sum_{i=1}^{A-1}
      \displaystyle\frac{z_{i}\,m_{\rm p}}{m_{Ai}}\, e^{-i\, \vb{k_{\rm ph}} \rhobf_{Ai}}\, 
      \bigl( \vb{e}^{(\alpha)} \cdot \vb{\tilde{p}}_{Ai} \bigr)  \\
  & - & 
    \displaystyle\frac{m_{\rm p}}{m_{A}}\, e^{i\, c_{\rm p}\, \vb{k_{\rm ph}} \vb{r}}\,
      \displaystyle\sum_{i=1}^{A} z_{i}\, e^{-i\, \vb{k_{\rm ph}} \rhobf_{Ai}}\, 
      \Bigl( \vb{e}^{(\alpha)} \cdot
        \displaystyle\sum_{k=1}^{A-1} \vb{\tilde{p}}_{Ak}
      \Bigr)
  \biggr\},
\end{array}
\label{eq.app.OpEmission.pA.4}
\end{equation}
\begin{equation}
\begin{array}{lll}
\vspace{0.4mm}
  \Delta \hat{H}_{\gamma M} & = &
  -\, i\, \mu_{0}\, \sqrt{\displaystyle\frac{2\pi c^{2}}{\hbar w_{\rm ph}}}\;
    e^{-i\, \vb{k_{\rm ph}} \vb{R}}\,
  \displaystyle\sum\limits_{\alpha=1,2}
  \biggl\{
    e^{i\, \vb{k_{\rm ph}} c_{\rm p}\, \vb{r}}\,
    \displaystyle\sum_{i=1}^{A-1}
      \mu_{i}\, e^{-i\, \vb{k_{\rm ph}} \rhobf_{Ai}}\, 
      \Bigl( \hat{\sigmabf}_{\rm p} \cdot \bigl[ \vb{\tilde{p}}_{Ai} \times \vb{e}^{(\alpha)} \bigr] \Bigr)  \\
  & - &
    e^{i\, \vb{k_{\rm ph}} c_{\rm p}\, \vb{r}}\,
    \displaystyle\sum_{i=1}^{A}
      \mu_{i}\, \displaystyle\frac{m_{Ai}}{m_{A}}\,
      e^{-i\, \vb{k_{\rm ph}} \rhobf_{Ai}}\,
      \displaystyle\sum_{k=1}^{A-1} \Bigl( \hat{\sigmabf}_{k} \cdot \bigl[ \vb{\tilde{p}}_{Ak} \times \vb{e}^{(\alpha)} \bigl] \Bigr)
  \biggr\}.
\end{array}
\label{eq.app.OpEmission.pA.5}
\end{equation}
The background term is given by
(see Eq.~(19) for scattering of $\alpha$ particles on nuclei in
Ref.~\cite{Liu_Maydanyuk_Zhang_Liu.2019.PRC.hypernuclei}),
\begin{equation}
\begin{array}{lcl}
  \hat{H}_{k} & = &
  i\, \hbar\, \mu_{0}\,
  \sqrt{\displaystyle\frac{2\pi c^{2}}{\hbar w_{\rm ph}}}\:
  e^{-i\, \vb{k_{\rm ph}} \vb{R}}\,
  \displaystyle\sum\limits_{\alpha=1,2}
    \biggl(
    \Bigl\{
      e^{-i\, c_{A}\, \vb{k_{\rm ph}} \vb{r}}\, \mu_{\rm p}\, \hat{\sigmabf}_{\rm p} +
      e^{i\, c_{\rm p}\, \vb{k_{\rm ph}} \vb{r}}\,
      \displaystyle\sum_{i=1}^{A}
        \mu_{i}\, e^{-i\, \vb{k_{\rm ph}} \rhobf_{Ai}}\, \hat{\sigmabf}_{i}
    \Bigr\} \cdot \bigl[ \vb{k_{\rm ph}} \times \vb{e}^{(\alpha)} \bigr]
    \biggl).
\end{array}
\label{eq.app.OpEmission.pA.6}
\end{equation}
%
Here,
$c_{A} = \frac{m_{A}}{m_{A}+m_{\rm p}}$,
$c_{\rm p} = \frac{m_{\rm p}}{m_{A} + m_{\rm p}}$,
and
$m_{\rm p}$ is the proton mass,
$m_{Ai}$ and $z_{i}$ are respectively the mass and electric charge of $i$-th nucleon,
$m_{A}$ is the target nucleus mass with the mass number $A$,
and
$\mu_{i}$ are the magnetic moments of proton or neutron in the nucleus in units of nuclear magneton $\mu_{0}$.
%
$\vb{\hat{P}}$, $\vb{\hat{p}}$ and $\vb{\tilde{p}}_{Ai}$ are the momenta corresponding
to $\vb{R}$, $\vb{r}$, $\rhobf_{Ai}$ for  $i=1 \ldots A-1$, respectively
(defined as $\vu{p}_{i} = -i\hbar\, \vb{d}/\vb{dr}_{i}$).
$\vb{e}^{(1)}$ and $\vb{e}^{(2)}$ are the polarization vectors of the photon with wave vector
$\vb{k}_{\rm ph}$,
and $w_{\rm ph} = k_{\rm ph} c = \bigl| \vb{k}_{\rm ph}\bigr|c$.
Note that $\vb{e}^{(\alpha)}$ ($\alpha=1,2$) are perpendicular to $\vb{k}_{\rm ph}$ in the Coulomb
gauge, satisfying
Eq.~(8) in Ref.~\cite{Liu_Maydanyuk_Zhang_Liu.2019.PRC.hypernuclei}.

\subsection{Matrix elements of emission of bremsstrahlung photons
\label{sec.13}}

We define the wave function of the proton-nucleus scattering as
\begin{equation}
  \Psi =
  \Phi (\vb{R}) \;
  \Phi_{pA} (\vb{r}) \;
  \psi_{A} (\beta_{A}),
\label{eq.app.2.6.1}
\end{equation}
where
\begin{equation}
\begin{array}{lcl}
  \psi_{\rm nucl} (\beta_{A}) =
  \psi_{\rm nucl} (1 \cdots A ) =
  \displaystyle\frac{1}{\sqrt{A!}}
  \displaystyle\sum\limits_{p_{A}}
    (-1)^{\varepsilon_{p_{A}}}\;
    \psi_{\lambda_{1}}(1)\;
    \psi_{\lambda_{2}}(2) \ldots
    \psi_{\lambda_{A}}(A),
\end{array}
\label{eq.app.2.6.2}
\end{equation}
following the formalism in Ref.~\cite{Maydanyuk_Zhang.2015.PRC}
for the proton-nucleus scattering (see Sect.~II.B, Eqs.~(10)--(13)),
and we add the description of many-nucleon structure of the nucleus
as in Ref.~\cite{Maydanyuk_Zhang_Zou.2016.PRC}.
Here,
$\beta_{A}$ is the set of numbers $1 \cdots A$ of nucleons in the nucleus $A$,
$\Phi (\vb{R})$ is the function describing the center-of-mass (CM) motion of the full nuclear
system in the laboratory frame,
$\Phi_{pA} (\vb{r})$ is the function describing the relative motion between
the scattered proton and target nucleus
(without the description of internal relative motion of nucleons in nucleus),
$\psi_{\rm nucl} (\beta_{A})$ is the many-nucleon wave function of the nucleus,
defined in Eq.~(12) Ref.~\cite{Maydanyuk_Zhang.2015.PRC}
on the basis of the one-nucleon wave function $\psi_{\lambda_{s}}(s)$,
and $\beta_{A}$ is the set of numbers $1 \cdots A$ of nucleons in the nucleus.
Summation in Eq.~(\ref{eq.app.2.6.2}) is performed over all $A!$
permutations of coordinates or states of nucleons.
One-nucleon wave function $\psi_{\lambda_{s}}(s)$ represents
the multiplication of space and spin-isospin
functions as
$\psi_{\lambda_{s}} (s) = \varphi_{n_{s}} (\vb{r}_{s})\, \bigl|\, \sigma^{(s)} \tau^{(s)}
\bigr\rangle$,
where
$\varphi_{n_{s}}$ is the space wave function of the nucleon with the index $s$,
$n_{s}$ is the number of state of the space wave function of the $s$-nucleon,
$\bigl|\, \sigma^{(s)} \tau^{(s)} \bigr\rangle$ is the spin-isospin wave
function of the $s$-nucleon.

We define the matrix element of the photon emission, using the wave functions
$\Psi_{i}$ and $\Psi_{f}$ of the full nuclear system, before the emission of photons
($i$-state) and after the emission ($f$-state), as
\begin{equation}
  F = \langle \Psi_{f} |\, \hat{H}_{\gamma} |\, \Psi_{i} \rangle.
\label{eq.13.1.1.p-nucl}
\end{equation}
In this matrix element we should integrate over all independent variables, i.e.
space variables $\vb{R}$, $\vb{r}$, $\rhobf_{Ai}$.
Furthermore, we should take into account the space representation of all the momenta used,
$\vu{P}$, $\vu{p}$, $\vb{\tilde{p}}_{Ai}$
(as
$\vu{P} = -i\hbar\, \vb{d/dR}$,
$\vu{p} = -i\hbar\, \vb{d/dr}$,
$\vb{\tilde{p}}_{A i} = -i\hbar\, \vb{d/d} \rhobf_{Ai}$).
Using formulas (\ref{eq.app.OpEmission.pA.1})--(\ref{eq.app.OpEmission.pA.6})
for the operator of photon emission, we calculate the full matrix element
of the bremsstrahlung photon emission.
The calculations are straightforward, and we write down the results
(see Ref.~\cite{Maydanyuk.2023.PRC.delta.Supplemental}, for details and references),
\begin{equation}
\begin{array}{lll}
  \langle \Psi_{f} |\, \hat{H}_{\gamma} |\, \Psi_{i} \rangle \;\; = \;\;
  \sqrt{\displaystyle\frac{2\pi\, c^{2}}{\hbar w_{\rm ph}}}\,
  M_{\rm full}, &
  \hspace{4ex} M_{\rm full} = M_{P} + M_{p}^{(E)} + M_{p}^{(M)} + M_{k} + M_{\Delta E} + M_{\Delta M},
\end{array}
\label{eq.13.1.2.p-nucl}
\end{equation}
where
\begin{equation}
\begin{array}{lllll}
\vspace{-0.1mm}
  M_{P} & = &
  \displaystyle\frac{\hbar\, (2\pi)^{3}}{m_{A} + m_{p}}\, \mu_{0}\,
  \displaystyle\sum\limits_{\alpha=1,2}
  \displaystyle\int\limits_{}^{}
    \Phi_{\rm p - nucl, f}^{*} (\vb{r})\;
  \biggl\{
    2\, m_{\rm p}\;
    \Bigl[
      e^{-i\, c_{A}\, \vb{k_{\rm ph}} \vb{r}} F_{p,\, {\rm el}} + e^{i\, c_{p}\, \vb{k_{\rm ph}} \vb{r}} F_{A,\, {\rm el}}
    \Bigr]\, \bigl( \vb{e}^{(\alpha)} \cdot \vb{K}_{i} \bigr)\; \\
  & + &
    i\: 
    \biggl( \Bigl[
      e^{-i\, c_{A}\, \vb{k_{\rm ph}} \vb{r}}\, \vb{F}_{p,\, {\rm mag}} + e^{i\, c_{p}\, \vb{k_{\rm ph}} \vb{r}}\, \vb{F}_{A,\, {\rm mag}}
    \Bigr] \cdot
    \bigl[ \vb{K}_{i} \cp \vb{e}^{(\alpha)} \bigr]
    \biggr)
  \biggr\}\;
  \Phi_{\rm p - nucl, i} (\vb{r})\; \vb{dr},
\end{array}
\label{eq.13.1.3.p-nucl}
\end{equation}
\begin{equation}
\begin{array}{lll}
\vspace{-0.2mm}
  M_{p}^{(E)} & = &
  2\,i \hbar\, (2\pi)^{3} \displaystyle\frac{m_{\rm p}}{\mu}\: \mu_{0}
  \displaystyle\sum\limits_{\alpha=1,2}
  \displaystyle\int\limits_{}^{}
    \Phi_{\rm p - nucl, f}^{*} (\vb{r})\;
    e^{-i\, \vb{k}_{\rm ph} \vb{r}}\;
    Z_{\rm eff} (\vb{k}_{\rm ph}, \vb{r}) \; 
    \Bigl( \vb{e}^{(\alpha)} \cdot \vb{\displaystyle\frac{d}{dr}}\; \Phi_{\rm p - nucl, i} (\vb{r}) \Bigr)\; \vb{dr}, \\
  M_{p}^{(M)} & = &
  -\, \hbar\, (2\pi)^{3} \displaystyle\frac{m_{\rm p}}{\mu}\: \mu_{0}
  \displaystyle\sum\limits_{\alpha=1,2}
  \displaystyle\int\limits_{}^{}
    \Phi_{\rm p - nucl, f}^{*} (\vb{r})\;
    e^{-i\, \vb{k}_{\rm ph} \vb{r}} \;
    \Bigl( \vb{M}_{\rm eff} (\vb{k}_{\rm ph}, \vb{r}) \cdot \Bigl[ \vb{\displaystyle\frac{d}{dr}} \times \vb{e}^{(\alpha)} \Bigr]\;
    \Phi_{\rm p - nucl, i} (\vb{r}) \Bigr) \; \vb{dr},
\end{array}
\label{eq.13.1.4.p-nucl}
\end{equation}
\begin{equation}
\begin{array}{lll}
  M_{\Delta E} & = &
  -\, (2\pi)^{3}\, 2\, \mu_{0}
  \displaystyle\sum\limits_{\alpha=1,2} 
  \displaystyle\int\limits_{}^{}
    \Phi_{\rm p - nucl, f}^{*} (\vb{r})\;
  \biggl\{
    e^{i\, c_{p}\, \vb{k_{\rm ph}} \vb{r}}\, 
    \Bigl( \vb{e}^{(\alpha)} \cdot \vb{D}_{A 1,\, {\rm el}} \Bigr) -
    \displaystyle\frac{m_{\rm p}}{m_{A}}\, e^{i\, c_{p}\, \vb{k_{\rm ph}} \vb{r}}\, 
     \Bigl( \vb{e}^{(\alpha)} \cdot \vb{D}_{A 2,\, {\rm el}} \Bigr)
  \biggr\}\;
  \Phi_{\rm p - nucl, i} (\vb{r})\; \vb{dr},
\end{array}
\label{eq.13.1.5.p-nucl}
\end{equation}
\begin{equation}
\begin{array}{lll}
  M_{\Delta M} & = &
  -\, i\, (2\pi)^{3}\,  \mu_{0}\,
  \displaystyle\sum\limits_{\alpha=1,2}
  \displaystyle\int\limits_{}^{}
    \Phi_{\rm p - nucl, f}^{*} (\vb{r})\;
  \biggl\{
    e^{i\, c_{p}\, \vb{k_{\rm ph}} \vb{r}}\; D_{A 1,\, {\rm mag}} (\vb{e}^{(\alpha)}) -
    e^{i\, c_{p}\, \vb{k_{\rm ph}} \vb{r}}\; D_{A 2,\, {\rm mag}} (\vb{e}^{(\alpha)})
  \biggr\}\;
  \Phi_{\rm p - nucl, i} (\vb{r})\; \vb{dr},
\end{array}
\label{eq.13.1.6.p-nucl}
\end{equation}
\begin{equation}
\begin{array}{lcl}
  M_{k} & = &
  i\, \hbar\, (2\pi)^{3}  \mu_{0}\,
  \displaystyle\sum\limits_{\alpha=1,2}
  \displaystyle\int\limits_{}^{}
    \Phi_{\rm p - nucl, f}^{*} (\vb{r})\;
    \Bigl(
      \bigl[ \vb{k_{\rm ph}} \cp \vb{e}^{(\alpha)} \bigr] \cdot
      \Bigl\{ e^{-i\, c_{A}\, \vb{k_{\rm ph}} \vb{r}}\, \vb{D}_{p,\, {\rm k}} + e^{i\, c_{p}\, \vb{k_{\rm ph}} \vb{r}}\, \vb{D}_{A,\, {\rm k}} \Bigr\}
    \Bigr)\;
    \Phi_{\rm p - nucl, i} (\vb{r})\; \vb{dr}.
\end{array}
\label{eq.13.1.7.p-nucl}
\end{equation}
Here, $\vb{K}_{i} = \vb{K}_{f} + \vb{k}_{\rm ph}$, and
$\mu = m_{\rm p} m_{A} / (m_{\rm p} + m_{A})$ the reduced mass, and
the effective electric charge and magnetic moment are
(see Eqs.~(28) and (29) in Ref.~
\cite{Maydanyuk.2023.PRC.delta.Supplemental}),
\begin{equation}
\begin{array}{lll}
\vspace{1.0mm}
  Z_{\rm eff} (\vb{k}_{\rm ph}, \vb{r}) =
  e^{i\, \vb{k_{\rm ph}} \vb{r}}\,
  \Bigl[
    e^{-i\, c_{A} \vb{k_{\rm ph}} \vb{r}}\, \displaystyle\frac{m_{A}}{m_{p} + m_{A}}\, F_{p,\, {\rm el}} -
    e^{i\, c_{p} \vb{k_{\rm ph}} \vb{r}}\, \displaystyle\frac{m_{p}}{m_{p} + m_{A}}\, F_{A,\, {\rm el}}
  \Bigr], \\
  \vb{M}_{\rm eff} (\vb{k}_{\rm ph}, \vb{r}) =
  e^{i\, \vb{k_{\rm ph}} \vb{r}}\,
  \Bigl[
    e^{-i\, c_{A} \vb{k_{\rm ph}} \vb{r}}\,  \displaystyle\frac{m_{A}}{m_{p} + m_{A}}\, \vb{F}_{p,\, {\rm mag}} -
    e^{i\, c_{p} \vb{k_{\rm ph}} \vb{r}}\,  \displaystyle\frac{m_{p}}{m_{p} + m_{A}}\, \vb{F}_{A,\, {\rm mag}}
  \Bigr].
\end{array}
\label{eq.13.1.8.p-nucl}
\end{equation}
$F_{{\rm p},\, {\rm el}}$,
$F_{A,\, {\rm el}}$,
$\vb{F}_{{\rm p},\, {\rm mag}}$,
$\vb{F}_{A,\, {\rm mag}}$,
$\vb{D}_{A 1,\, {\rm el}}$,
$\vb{D}_{A 2,\, {\rm el}}$,
$D_{A 1,\, {\rm mag}}$,
$D_{A 2,\, {\rm mag}}$,
$\vb{D}_{{\rm p},\, {\rm k}}$,
$\vb{D}_{A,\, {\rm k}}$,
$D_{{\rm p}, P\, {\rm el}}$,
$D_{A,P\, {\rm el}}$,
$\vb{D}_{{\rm p}, P\, {\rm mag}}$,
$\vb{D}_{A,P\, {\rm mag}}$
are the electric and magnetic form factors 
of proton and nucleus defined in Sec.~\ref{sec.app.form_factors} 
[see Eqs.~(\ref{eq.app.form_factors.1.1})--(\ref{eq.app.form_factors.1.5})].
The proton form factors $F_{{\rm p},\, {\rm el}}$ and $D_{{\rm p}, P\, {\rm el}}$
depend on the proton charge but they are constant.
The form factors $\vb{F}_{{\rm p},\, {\rm mag}}$ and $\vb{D}_{{\rm p}, P\, {\rm mag}}$
depend on the proton magnetic moment.

\subsection{Matrix elements in multipole expansion
\label{sec.14}}

We use monopole approximation of the effective electric charge and magnetic moment
(\ref{eq.13.1.8.p-nucl})
(see Eqs.~(40) and (49) in Ref.~\cite{Maydanyuk.2023.PRC.delta.Supplemental}) as,
\begin{equation}
\begin{array}{llll}
  Z_{\rm eff}^{\rm (mon)} (\vb{k}_{\rm ph}) = \displaystyle\frac{m_{A}\, z_{p} - m_{\rm p}\, Z_{\rm A}(\vb{k}_{\rm ph})}{m_{A} + m_{\rm p}}, \\
  \vb{M}_{\rm eff}^{\rm (mon)} (\vb{k}_{\rm ph}) =
  \displaystyle\frac{\mu}{m_{\rm p}}\, \Bigl[ \vb{F}_{A,\, {\rm mag}} (\vb{k}_{\rm ph}) - \vb{F}_{B,\, {\rm mag}} (\vb{k}_{\rm ph}) \Bigr] =
  -\, \displaystyle\frac{\mu}{m_{\rm p}}\,
  \alpha \; (\vb{e}_{\rm x} + \vb{e}_{\rm z}).
\end{array}
\label{eq.multimple.1}
\end{equation}
%
Here
\begin{equation}
\begin{array}{lll}
  \alpha & = &
    \displaystyle\frac{m_{\rm p}}{m_{A}}\, Z_{\rm A} (\vb{k}_{\rm ph})\, \bar{\mu}_{\rm pn}^{\rm (A)} -
    z_{\rm p}\, \mu_{\rm p},
\end{array}
\label{eq.resultingformulas.6}
\end{equation}
where $Z_{A} (\vb{k}_{\rm ph})$ is electric form factor of the nucleus
defined in Eqs.~(\ref{eq.app.form_factors.1.1}).
%
The calculation of integrals in the matrix elements
(\ref{eq.13.1.2.p-nucl})--(\ref{eq.13.1.7.p-nucl}) is straightforward
when we use the multipole expansion of the wave function of the photons~\cite{Eisenberg_Greiner.1970book,Greiner-book1996,Maydanyuk03}
(see Eqs.~(22)--(28) in Ref.~\cite{Maydanyuk.2023.PRC.delta} for details).
As a result, we obtain the solutions for coherent terms as
\begin{equation}
\begin{array}{lll}
\vspace{1.0mm}
  M_{p}^{(E,\, {\rm mon},\, 0)} =
  -\, \hbar\, (2\pi)^{3}\,
  \displaystyle\frac{\mu_{0}}{\sqrt{3}} \;
  \displaystyle\frac{m_{\rm p}\, Z_{\rm eff}^{\rm (mon,\, 0)}}{\mu} \;
  \Bigl(
    J_{1}(0,1,0) -
    \displaystyle\frac{47}{40} \sqrt{\displaystyle\frac{1}{2}} \; J_{1}(0,1,2)
  \Bigr), \\
  M_{p}^{(E,\, {\rm mon},\, 0)} + M_{p}^{(M,\, {\rm mon},\, 0)} =
   M_{p}^{(E,\, {\rm mon},\, 0)} \;
  \Bigl(
    1 -
    i\: \displaystyle\frac{\mu } {2\, m_{\rm p}\, Z_{\rm eff}^{\rm (mon,\, 0)}}\; \alpha
  \Bigr), \\
\end{array}
\label{eq.multimple.2}
\end{equation}
and solutions for incoherent and background terms as
\begin{equation}
\begin{array}{llllll}
\vspace{1.5mm}
  M_{\Delta E} = 0, \\
\vspace{1.5mm}
  M_{\Delta M} =
  \hbar\, (2\pi)^{3}\,
  \displaystyle\frac{\sqrt{3}}{2}\,
  \mu_{0}\, k_{\rm ph} \;
  \displaystyle\frac{A-1}{2A}\: \bar{\mu}_{\rm pn}^{\rm (A)}
  Z_{\rm A} (\vb{k}_{\rm ph}) \; \tilde{J}\, (+c_{\rm p}, 0,1,1), \\
\vspace{1.5mm}
  M_{k} =
  -\, \hbar\, (2\pi)^{3}\,
  \displaystyle\frac{\sqrt{3}}{2}\:
  \mu_{0}\, k_{\rm ph}\,
  \Bigl\{
    \mu_{\rm p} \; z_{\rm p} \; \tilde{J}\, (- c_{A}, 0,1,1) +
    \bar{\mu}_{\rm pn}^{(A)} \; Z_{\rm A} (\vb{k}_{\rm ph}) \; \tilde{J}\, (+ c_{\rm p}, 0,1,1)
  \Bigr\}, \\
  M_{\Delta M} + M_{k} =
  -\, \hbar\, (2\pi)^{3}\,
  \displaystyle\frac{\sqrt{3}}{2}\,
  \mu_{0}\, k_{\rm ph}\,
  \Bigl\{
    \mu_{\rm p} \; z_{\rm p} (\vb{k}_{\rm ph}) \; \tilde{J}\, (- c_{A}, 0,1,1) +
    \displaystyle\frac{A+1}{2A}\, \bar{\mu}_{\rm pn}^{(A)} \; Z_{\rm A} (\vb{k}_{\rm ph}) \;
    \tilde{J}\, (+c_{\rm p}, 0,1,1)
  \Bigr\}.
\end{array}
\label{eq.app.model.simplestcase.result.6}
\end{equation}
Here,
$\bar{\mu}_{\rm pn}^{\rm (A)} = \mu_{\rm p} + \kappa_{A}\,\mu_{\rm n}$,
$\kappa_{A} = N_{A}/Z_{A}$,
$\mu_{\rm p}$ and $\mu_{\rm n}$ are the proton and neutron magnetic moments, respectively,
and $A$ and $N_{A}$ are the numbers of nucleons and neutrons
of the nucleus.
We have
\begin{equation}
\begin{array}{lll}
  \alpha & = &
  \displaystyle\frac{m_{\rm p}}{\mu}\;
  \Bigl[
    \displaystyle\frac{m_{\rm p}}{m_{A}}\, Z_{\rm A} (\vb{k}_{\rm ph})\, \bar{\mu}_{\rm pn}^{\rm(A)}
    \displaystyle\frac{m_{\rm p}}{m_{B}}\, Z_{\rm B} (\vb{k}_{\rm ph})\, \bar{\mu}_{\rm pn}^{\rm(B)}
 \Bigr]
 \\
 & = &
  \displaystyle\frac{m_{\rm p}^{2}}{\mu^{2}}\;
  \Bigl[
    \displaystyle\frac{m_{B}\, Z_{\rm A} (\vb{k}_{\rm ph})\, \bar{\mu}_{\rm pn}^{\rm (A)} - m_{A}\, Z_{\rm B} (\vb{k}_{\rm ph})\, \bar{\mu}_{\rm pn}^{\rm (A)}}{m_{A} + m_{B}}
  \Bigr] \ne
  \displaystyle\frac{m_{\rm p}^{2}}{\mu^{2}}\,
  Z_{\rm eff}^{\rm (mon)} (\vb{k}_{\rm ph})\,
  \bar{\mu}_{\rm pn}^{\rm (A)}.
\end{array}
\label{eq.multimple.3}
\end{equation}

The radial integrals are
\begin{equation}
\begin{array}{llllll}
\hspace{-2ex}
  J_{1}(l_{i},l_{f},n) & = & \displaystyle\int\limits^{+\infty}_{0} \displaystyle\frac{dR_{i}(r,
l_{i})}{dr}\: R^{*}_{f}(l_{f},r)\, j_{n}(k_{\rm ph}r)\; r^{2} dr, &
\hspace{2ex}
  \tilde{J}\,(c, l_{i},l_{f},n) & = & \displaystyle\int\limits^{+\infty}_{0} R_{i}(l_{i},
r)\, R^{*}_{f}
  (l_{f},r)\, j_{n}(c\, k_{\rm ph}r)\; r^{2} dr.
\end{array}
\label{eq.multimple.4}
\end{equation}
Here, $R_{i,f}$ is the radial part of the wave function $\Phi_{A - B} (\vb{r})$ in $i$-state
or $f$-state, and
$j_{\rm n}(k_{\rm ph}r)$ is spherical Bessel function of order $n$.

We define the cross sections of the bremsstrahlung photon emission
in the proton-nucleus scattering
on the basis of the full matrix element $p_{fi}$
in the frameworks of formalism given in
Refs.~\cite{Maydanyuk.2023.PRC.delta,Maydanyuk_Zhang_Zou.2016.PRC,
Maydanyuk.2012.PRC,Maydanyuk_Zhang.2015.PRC}
(see Eq.~(29) in Ref.~\cite{Maydanyuk.2023.PRC.delta} and reference therein),
and we do not repeat it here. Finally, we obtain the bremsstrahlung cross sections as
\begin{equation}
\begin{array}{llll}
  \displaystyle\frac{d \sigma}{dw_{\rm ph}} =
    \displaystyle\frac{e^{2}}{2\pi\,c^{5}}\: \displaystyle\frac{w_{\rm ph}\,E_{i}}{m_{\rm p}^{2}\,k_{i}}\:
    \bigl| p_{fi} \bigr|^{2},
    \hspace{3ex} &
  \displaystyle\frac{d^{2} \sigma}{dw_{\rm ph}\, d \cos \theta} =
    \displaystyle\frac{e^{2}}{2\pi\,c^{5}}\: \displaystyle\frac{w_{\rm ph}\,E_{i}}{m_{\rm p}^{2}\,k_{i}}\:
    \bigl\{ p_{fi} \displaystyle\frac{d\,p_{fi}^{*}}{d\, \cos \theta}  +
    {\rm c.\,c.}\bigr\},
    \hspace{3ex} &
  M_{\rm full} = - \displaystyle\frac{e}{m_{\rm p}}\, p_{fi},
\end{array}
\label{eq.model.bremprobability.1}
\end{equation}
where c.\,c.~is complex conjugation.
We also calculate the different contributions of the emitted photons
to the full bremsstrahlung spectrum
on the basis of the associated matrix elements in Eqs.~(\ref{eq.app.model.simplestcase.result.6}).
The matrix elements are calculated based on the wave functions with quantum numbers $l_{i}=0$,
$l_{f}=1$ and $l_{\rm ph}=1$,
where, $l_{i}$ and $l_{f}$ are the orbital quantum numbers of the wave
function $\Phi_{pA} (\vb{r})$ defined in Eq.~(\ref{eq.app.2.6.1})
for the states before and after the photon emission, respectively,
$l_{\rm ph}$ is the orbital quantum number of photon in the multipole approach.



\section{Form factors of nuclei
\label{sec.app.form_factors}}

In this section we present the form factors used in the calculations of matrix elements in
Eqs.~(\ref{eq.13.1.3.p-nucl})--(\ref{eq.13.1.8.p-nucl})
(details of the calculation of the form factors for the proton-nucleus scattering are given in
Ref.~\cite{Maydanyuk.2023.PRC.delta}, and in Sec.~I of Supplemental Material
of that paper~\cite{Maydanyuk.2023.PRC.delta.Supplemental}, Eqs.~(16), (21), (23) and (25)).
Here, we write down the solutions for \emph{electric and magnetic form factors}
for the nucleus with the nucleon number $A$ as
\begin{equation}
\begin{array}{lll}
\vspace{1mm}
  F_{A,\, {\rm el}} & = &
    \displaystyle\sum\limits_{i=1}^{A}
    \Bigl\langle \psi_{\rm nucl, f} (\beta_{A}) \Bigl|\,
      z_{i}\, e^{-i \vb{k}_{\rm ph} \rhobf_{A i} }
    \Bigr|\, \psi_{\rm nucl, i} (\beta_{A}) \Bigr\rangle =
  Z_{A} (\vb{k}_{\rm ph}), \\

  \vb{F}_{A,\, {\rm mag}} & = &
    \displaystyle\frac{1}{m_{A}}
    \displaystyle\sum_{i=1}^{A}
    \Bigl\langle \psi_{\rm nucl, f} (\beta_{A})\, \Bigl|\,
        \mu_{i}\, m_{Ai}\; e^{-i\, \vb{k_{\rm ph}} \rhobf_{Ai}}\, \hat{\sigmabf}_{i}
    \Bigr| \psi_{\rm nucl, i} (\beta_{A}) \Bigr\rangle =
  \displaystyle\frac{m_{\rm p}}{m_{A}}\,
    Z_{\rm A} (\vb{k}_{\rm ph})\, \bar{\mu}_{\rm pn}^{\rm (A)} \;
    (\vb{e}_{\rm x} + \vb{e}_{\rm z}),
\end{array}
\label{eq.app.form_factors.1.1}
\end{equation}
%
\begin{equation}
\begin{array}{lll}
\vspace{1mm}
  D_{A,P\, {\rm el}} = &
    \displaystyle\sum\limits_{i=1}^{A}
    \Bigl\langle \psi_{\rm nucl, f} (\beta_{A}) \Bigl|\,
      z_{i}\, e^{-i \vb{k}_{\rm ph} \rhobf_{A i} }
    \Bigr|\, \psi_{\rm nucl, i} (\beta_{A}) \Bigr\rangle =
  F_{A,\, {\rm el}}, \\

  \vb{D}_{A,P\, {\rm mag}} = &
    \displaystyle\frac{1}{m_{A}}
    \displaystyle\sum_{i=1}^{A}
    \Bigl\langle \psi_{\rm nucl, f} (\beta_{A})\, \Bigl|\,
        \mu_{i}\, m_{A i}\; e^{-i\, \vb{k_{\rm ph}} \rhobf_{Ai}}\, \hat{\sigmabf}_{i}
    \Bigr| \psi_{\rm nucl, i} (\beta_{A}) \Bigr\rangle =
  \vb{F}_{A,\, {\rm mag}},
\end{array}
\label{eq.app.form_factors.1.2}
\end{equation}

\begin{equation}
\begin{array}{lll}
\vspace{1mm}
  \vb{D}_{A 1,\, {\rm el}} = &
    \displaystyle\sum\limits_{i=1}^{A-1}
      \displaystyle\frac{z_{i} m_{\rm p}}{m_{Ai}}\,
      \Bigl\langle \psi_{A, f} (\beta_{A})\, \Bigl|\,
        e^{-i \vb{k}_{\rm ph} \rhobf_{Ai}} \vb{\tilde{p}}_{Ai}
      \Bigr|\,  \psi_{A,i}  (\beta_{A})  \Bigr\rangle =

  \displaystyle\frac{\hbar}{2}\; \displaystyle\frac{A-1}{A}\; \vb{k}_{\rm ph}\; Z_{\rm A} (\vb{k}_{\rm ph}), \\

  \vb{D}_{A 2,\, {\rm el}} = &
    \displaystyle\sum\limits_{i=1}^{A}
      z_{i}\,
      \Bigl\langle \psi_{A, f} (\beta_{A})\, \Bigl|\,
        e^{-i \vb{k}_{\rm ph} \rhobf_{Ai}}
        \Bigl( \displaystyle\sum_{k=1}^{A-1} \vb{\tilde{p}}_{Ak} \Bigr)
      \Bigr|\,  \psi_{A,i} (\beta_{A}) \Bigr\rangle \sim \vb{k}_{\rm ph},
\end{array}
\label{eq.app.form_factors.1.3}
\end{equation}


\begin{equation}
\begin{array}{lll}
\vspace{1mm}
  \displaystyle\sum\limits_{\alpha=1,2} D_{A 1,\, {\rm mag}}^{(\alpha)} & = &
  \displaystyle\sum\limits_{\alpha=1,2}
    \displaystyle\sum\limits_{i=1}^{A-1}
    \mu_{i}\,
    \Bigl\langle \psi_{A, f} (\beta_{A})\, \Bigl|\,
      e^{-i\, \vb{k_{\rm ph}} \rhobf_{Ai}}\; 
      \Bigl( \hat{\sigmabf}_{i} \cdot \bigl[ \vb{\tilde{p}}_{Ai} \times \vb{e}^{(\alpha)} \bigr] \Bigr)
      \Bigr|\,  \psi_{A,i} (\beta_{A})  \Bigr\rangle = \\
\vspace{1mm}
  & = &
  -\, \displaystyle\frac{\hbar\, (A-1)}{2\,A}\; \bar{\mu}_{\rm pn}\, k_{\rm ph} \; Z_{\rm A} (\vb{k}_{\rm ph}), \\ 
  D_{A 2,\, {\rm mag}}^{(\alpha)} & = &
    \displaystyle\sum\limits_{i=1}^{A}
      \mu_{i}\,
      \displaystyle\frac{m_{Ai}}{m_{A}}\,
      \Bigl\langle \psi_{A, f} (\beta_{A})\, \Bigl|\,
      e^{-i\, \vb{k_{\rm ph}} \rhobf_{Ai}}\; 
      \Bigl( \hat{\sigmabf}_{i} \cdot
        \Bigl[ \displaystyle\sum_{k=1}^{A-1} \vb{\tilde{p}}_{Ak} \times \vb{e}^{(\alpha)} \Bigr] \Bigr)
        \Bigr|\, \psi_{A, i} (\beta_{A}) \Bigr\rangle = 0,
\end{array}
\label{eq.app.form_factors.1.4}
\end{equation}

\begin{equation}
\begin{array}{lll}
  \vb{D}_{A,\, {\rm k}} = &
  \displaystyle\sum\limits_{i=1}^{A}
    \mu_{i}\,
    \Bigl\langle \psi_{A, f} (\beta_{A})\, \Bigl|\,
      e^{-i\, \vb{k_{\rm ph}} \rhobf_{Ai}}\, \hat{\sigmabf}_{i}
    \Bigr|\,  \psi_{A, i} (\beta_{A}) \Bigr\rangle =

  \bar{\mu}_{\rm pn}\; (\vb{e}_{\rm x} + \vb{e}_{\rm z}) \; Z_{\rm A} (\vb{k}_{\rm ph}).
\end{array}
\label{eq.app.form_factors.1.5}
\end{equation}
One can easily obtain the corresponding form factors for the proton from the above formulas
(for example, by replacing,
$Z_{A} (\vb{k}_{\rm ph}) = Z_{A} \to z_{\rm p}$).
Calculations of $Z_{\rm A} (\vb{k}_{\rm ph})$ for 
nuclei with different masses are given in Appendix~A of
Ref.~\cite{Maydanyuk_Zhang_Zou.2016.PRC}.
(See Eq.~(A9) for $\alpha$ particle,
$Z_{\rm A} (\vb{k}_{\rm ph})$, and for heavier nuclei, they are obtained from
Eqs.~(A18)--(A20) applying a limit of $\rho_{i} \to 0$.
For nuclei, in the first approximation, one can use $Z_{\rm A} (\vb{k}_{\rm ph}) = Z_{\rm A}$,
where $Z_{\rm A}$ is the electric charge of the nucleus.)



\begin{thebibliography}{99}

\bibitem{Maydanyuk.2023.PRC.delta}
  S.~P.~Maydanyuk,
\newblock
  \emph{Enhancement of incoherent bremsstrahlung in proton-nucleus scattering in the $\Delta$-resonance energy region},
\newblock
  Phys. Rev. \textbf{C 107}, 024618 (2023).

  
\bibitem{Maydanyuk_Zhang.2015.PRC}
  S.~P.~Maydanyuk and P.-M.~Zhang,
\newblock
  \emph{New approach to determine proton-nucleus interactions from experimental bremsstrahlung data},
\newblock
  Phys. Rev. \textbf{C 91}, 024605 (2015).


\bibitem{Goethem.2002.PRL}
  M.~J.~van~Goethem, L.~Aphecetche, J.~C.~S.~Bacelar, H.~Delagrange, J.~Diaz, D.~d'Enterria, M.~Hoefman,
  R.~Holzmann, H.~Huisman, N.~Kalantar-Nayestanaki, A.~Kugler, H.~L\"{o}hner, G.~Martinez,
  J.~G.~Messchendorp, R.~W.~Ostendorf, S.~Schadmand, R.~H.~Siemssen, R.~S.~Simon,
  Y.~Schutz, R.~Turrisi, M.~Volkerts, V.~Wagner, and H.~W.~Wilschut,
\newblock
  \textit{Suppresion of soft nuclear bremsstrahlung in proton-nucleus collisions},
\newblock
  Phys. Rev. Lett. \textbf{88}, 122302 (2002).


\bibitem{Clayton.1992.PRC}
  J.~Clayton, W.~Benenson, M.~Cronqvist, R.~Fox, D.~Krofcheck, R.~Pfaff,
  M.~F.~Mohar,
  C.~Bloch, D.~E.~Fields,
  \newblock
  \textit{High energy gamma ray production in proton-induced reactions at 104, 145, and 195 MeV},
\newblock
Phys. Rev. \textbf{C 45}, 1815 (1992).


\bibitem{Clayton.1991.PhD}
  J.~E.~Clayton,
\newblock
  \emph{High energy gamma ray production in proton induced reactions
  at energies of 104, 145, and 195 MeV},
\newblock
  PhD thesis (Michigan State University, 1991).


\bibitem{Liu_Maydanyuk_Zhang_Liu.2019.PRC.hypernuclei}
  X.~Liu, S.~P.~Maydanyuk, P.-M.~Zhang, and L.~Liu,
\newblock
  \emph{First investigation of hypernuclei in reactions via analysis of emitted bremsstrahlung photons},
\newblock
  Phys. Rev. \textbf{C 99}, 064614 (2019).


\bibitem{Guichon:1987jp}
  P.~A.~M.~Guichon,
  {\it A Possible Quark Mechanism for the Saturation of Nuclear Matter},
  Phys.\ Lett.\ B {\bf 200}, 235 (1988).


\bibitem{Saito:2005rv}
  K.~Saito, K.~Tsushima and A.~W.~Thomas,
  {\it Nucleon and hadron structure changes in the nuclear medium and impact on observables},
  Prog.\ Part.\ Nucl.\ Phys.\  {\bf 58}, 1 (2007).


\bibitem{Krein:2017usp}
  G.~Krein, A.~W.~Thomas and K.~Tsushima,
  {\it Nuclear-bound quarkonia and heavy-flavor hadrons}, 
  Prog.\ Part.\ Nucl.\ Phys.\  {\bf 100}, 161 (2018).


\bibitem{Guichon:2018uew}
  P.~A.~M.~Guichon, J.~R.~Stone and A.~W.~Thomas,
  {\it Quark-Meson-Coupling (QMC) model for finite nuclei, nuclear matter and beyond}, 
Prog. Part. Nucl. Phys. \textbf{100}, 262 (2018).


\bibitem{Lu:1997mu}
  D.~H.~Lu, A.~W.~Thomas, K.~Tsushima, A.~G.~Williams and K.~Saito,
  {\it In-medium electron-nucleon scattering},
Phys. Lett. B \textbf{417}, 217 (1998).


\bibitem{Lu:1998tn}
  D.~H.~Lu, K.~Tsushima, A.~W.~Thomas, A.~G.~Williams and K.~Saito,
  {\it Electromagnetic form-factors of the bound nucleon},
Phys. Rev. \textbf{C 60}, 068201 (1999).





\bibitem{Maydanyuk.2025.Supplemental}
See Supplemental Material 
for this paper
for details of calculations of
the operator of emission and matrix element of emission of photons.


\bibitem{Maydanyuk.2023.PRC.delta.Supplemental}
  See Supplemental Material at
{\color{blue}\href{https://journals.aps.org/prc/abstract/10.1103/PhysRevC.107.024618#supplemental}{https://doi.org/10.1103/PhysRevC.107.024618}}
for details of calculations of the matrix element of emission.




\bibitem{Maydanyuk_Zhang_Zou.2016.PRC}
  S.~P.~Maydanyuk, P.-M.~Zhang, and L.-P.~Zou,
\newblock
  \emph{New approach for obtaining information on the many-nucleon structure in $\alpha$ decay from accompanying bremsstrahlung emission},
\newblock
  Phys. Rev. \textbf{C 93}, 014617 (2016).

\bibitem{Maydanyuk.2012.PRC}
  S.~P.~Maydanyuk,
\newblock
  \emph{Model for bremsstrahlung emission accompanying interactions between protons and nuclei from low energies up to intermediate energies: Role of magnetic emission},
\newblock
  Phys. Rev. \textbf{C 86}, 014618 (2012).


\bibitem{Tsushima:2020gun}
  K.~Tsushima,
\newblock
  \emph{Magnetic moments of the octet, decuplet, low-lying charm, and low-lying bottom baryons in a nuclear medium},
\newblock
PTEP \textbf{043D02} (2022).

\end{thebibliography}

\begin{thebibliography}{99}
\bibitem{Maydanyuk_Zhang.2015.PRC}
  S.~P.~Maydanyuk and P.-M.~Zhang,
\newblock
  \emph{New approach to determine proton-nucleus interactions from experimental bremsstrahlung data},
\newblock
  Phys. Rev. \textbf{C 91}, 024605 (2015).




\bibitem{Maydanyuk_Zhang_Zou.2016.PRC}
  S.~P.~Maydanyuk, P.-M.~Zhang, and L.-P.~Zou,
\newblock
  \emph{New approach for obtaining information on the many-nucleon structure in $\alpha$ decay from accompanying bremsstrahlung emission},
\newblock
  Phys. Rev. \textbf{C 93}, 014617 (2016).

\bibitem{Maydanyuk.2012.PRC}
  S.~P.~Maydanyuk,
\newblock
  \emph{Model for bremsstrahlung emission accompanying interactions between protons and nuclei from low energies up to intermediate energies: Role of magnetic emission},
\newblock
  Phys. Rev. \textbf{C 86}, 014618 (2012).






\bibitem{Liu_Maydanyuk_Zhang_Liu.2019.PRC.hypernuclei}
  X.~Liu, S.~P.~Maydanyuk, P.-M.~Zhang, and L.~Liu,
\newblock
  \emph{First investigation of hypernuclei in reactions via analysis of emitted bremsstrahlung photons},
\newblock
  Phys. Rev. \textbf{C 99}, 064614 (2019).


  
  
\bibitem{Maydanyuk.2023.PRC.delta}
  S.~P.~Maydanyuk,
\newblock
  \emph{Enhancement of incoherent bremsstrahlung in proton-nucleus scattering in the $\Delta$-resonance energy region},
\newblock
  Phys. Rev. \textbf{C 107}, 024618 (2023).



  


\bibitem{Maydanyuk.2023.PRC.delta.Supplemental}
  See Supplemental Material at
{\color{blue}\href{https://journals.aps.org/prc/abstract/10.1103/PhysRevC.107.024618#supplemental}{https://doi.org/10.1103/PhysRevC.107.024618}}
for details of calculations of the matrix element of emission.




\bibitem{Eisenberg_Greiner.1970book}
  J.~M.~Eisenberg and W.~Greiner,
\newblock
  \emph{Nuclear Theory: Exitation mechanisms of the nucleus. Electromagnetic and weak interactions},
  Vol.~2
\newblock
  (North-Holland Publishing Company, Amsterdam - London, 1970). ---
%
  J.~M.~Eisenberg and W.~Greiner,
\newblock
  \emph{Mehanizmi vozbuzhdenia yadra. Elektromagnitnoe i slaboe vzaimodeistvia},
  Vol.~2
\newblock
  (Atomizdat, Moskva, 1973), p.~347 [in Russian].



\bibitem{Greiner-book1996}
W.~Greiner and  J.~A. Maruhn,
 \emph{Nuclear Models}, 
Springer, 1996,
ISBN 978-3-642-60980-1


\bibitem{Maydanyuk03}
S.~P.~Maydanyuk and V.~S.~Olkhovsky,
\emph{Does sub-barrier bremsstrahlung in $\alpha$-decay of $\isotope[210]{Pb}$ exist?}
Prog. Theor. Phys. \textbf{109}, 203 (2003).


\end{thebibliography}
\end{document}